\documentclass[twocolumn]{aastex6}
\usepackage{amsmath,amssymb,bm, graphicx,xspace,multirow,threeparttable,ulem}
\usepackage{epsfig}
\usepackage{color}
\RequirePackage{lineno}
\usepackage{txfonts}
\usepackage{float}

\def\link_col{blue}
\def\gray{$\gamma$-ray\xspace}

\begin{document}

\title{Morphology of gamma-ray halos around middle-aged pulsars: influence of the pulsar proper motion}

\author{Yi Zhang$^{1,2}$, Ruo-Yu Liu$^{1,2}$, S.~Z. Chen$^{3,4}$, Xiang-Yu Wang$^{1,2}$}
\affil{$^1$School of Astronomy and Space Science, Xianlin Road 163, Nanjing University, Nanjing
210023, China; \textcolor{blue}{ryliu@nju.edu.cn; xywang@nju.edu.cn}\\
$^2$Key laboratory of Modern Astronomy and Astrophysics (Nanjing University), Ministry of Education, Nanjing 210023, People's Republic of China\\
$^3$ Key Laboratory of Particle Astrophysics, Institute of High Energy Physics, Chinese Academy of Sciences, 100049 Beijing, China\\
$^4$TIANFU Cosmic Ray Research Center, Chengdu, Sichuan,  China
}

\begin{abstract}
Recently, \gray halos of a few degree extension have been detected around two middle-aged pulsars, namely, Geminga and PSR B0656+14, by the High Altitude Water Cherenkov observatory (HAWC). The \gray radiation arise from relativistic electrons that escape the pulsar wind nebula and diffuse in the surrounding medium. The diffusion coefficient is found to be significantly lower than the average value in the Galactic disk. If so, given a typical transverse velocity of $300-500{\,\rm km /s}$ for a pulsar, its displacement could be important in shaping the morphology of its \gray halos. Motivated by this, we study the morphology of pulsar halos considering the proper motion of pulsar.  We define three evolutionary phases of pulsar halo to categorize its morphological features. The morphology of pulsar halos below 10\,TeV is double peaked or single peaked with an extended tail, which depends on the electron injection history. Above 10 TeV, the morphology of  pulsar halos is nearly spherical, due to the short cooling timescale ($<50$\,kyr) for tens TeV electrons. We also quantitatively evaluate the separation between the pulsar and the center of the \gray halo, as well as the influence of different assumptions for the pulsar characteristics and the injected electrons.
Our results suggest that the separation between   the center of the \gray halo above 10\,TeV and the associated pulsar is usually too small to be observable by HAWC or LHAASO. Hence
, our results provide a useful approach to constrain the origin of  extended sources at very high energies.

\end{abstract}

\section{Introduction}

Pulsar wind nebulae (PWNe) are bubbles of relativistic electrons and positrons\footnote{hereafter we do not distinguish positrons from electrons for simplicity}, accelerated when a pulsar's relativistic wind interacts with its environment, either the supernova remnant (SNR) or the interstellar medium (ISM) \citep[e.g.][]{GS06}. Pulsar is formed in a supernova (SN) explosion that drives a strong blast wave with $\sim 10^{51}\,$erg kinetic energy expanding into ambient medium. The pulsar and its PWN are initially surrounded by the SNR. 
The SNR blast wave at first moves outward freely at a speed $>(5-10)\times10^3 {\,\rm km/s}$, while the asymmetry in the SN explosion would give the pulsar a natal velocity.
Observationally, we expect to see a rapidly expanding SNR with size of $\sim 1-10$\,pc, a reasonably symmetric PWN near its center  with a typical size at sub-pc to pc level, and a young pulsar at the center of the PWN, at early evolutionary epoch of the SNR-PWN system.

The expanding supernova shell starts to slow down as it sweeps up comparable mass of the surrounding ISM to that of the SN ejecta at a time $t_{\rm sd}\simeq 1400\,(M_{\rm ej}/10\,M_\odot)^{5/6}(E_{\rm SN}/10^{51}{\,\rm erg})^{-1/2}({n_{\rm ISM}}/{1{\,\rm cm^{-3}}})^{-1/3}$\,yr after the SN explosion \citep{Draine11}.
Because the SNR is decelerating, the pulsar ultimately penetrates and then escapes the shell at a time  $t_{\rm cross}\simeq 45\, ({E_{\rm SN}}/{10^{51}{\rm erg}})^{1/3} ({n_{\rm ISM}}/{1{\rm cm^{-3}}})^{-1/3} ({v_{\rm p}}/{400{\rm Km s^{-1}}})^{-5/3}\, {\rm kyr}$ \citep{Vanderswaluw03}.
After that, the pulsar proceeds to move through the ambient ISM with proper velocity $v_{\rm p}$ while the PWN at this stage is compact, smaller than 1\,pc, and filled with recently injected particles \citep{Kargaltsev13}.

The High-Altitude Water Cherenkov Observatory (HAWC) recently reported discovery of spatially extended TeV sources surrounding two middle-aged ($t_{\rm age}=100-400\, {\rm kyr}$) pulsars, namely, Geminga and PSR B0656+14 \citep[][see also \citealt{Milagro09}]{HAWC17_Geminga}. The intensity profile of the observed extended TeV sources can be explained with the inverse Compton (IC) scatterings of diffusing electrons, which are injected from the pulsars, on the cosmic microwave background (CMB) and the interstellar radiation field (ISRF) \citep{HAWC17_Geminga, Lopez18, Tang19, Liu19, Liu19_prl, Johan19, DiMauro19}. The physical extension of the sources is at least 30\,pc,  which is much larger than the size of the PWN of the corresponding pulsars, likely indicating that accelerated electrons escape the PWN and produce the halo-like emissions in the ambient ISM of the pulsars.

In addition to Geminga and PSR B0656+14, many more such pulsar halos \footnote{In some literature, they are also called TeV halos since they are initially discovered at the TeV band.} could have been already detected by instruments such as HAWC, The High Energy Stereoscopic System (HESS), Fermi-Large Area Telescope (Fermi-LAT) and LHAASO. The HAWC Collaboration has recently released the 2HWC catalog \citep{HAWC18} which contains 39 sources detected close to the Galactic plane. Some of them have an extended morphology, and are spatially close to powerful Galactic pulsars of middle ages. More recently, in the new released 3HWC catalog \citep{HAWC20_3HWC} they highlight 12 extended TeV sources as the potential pulsar halos. In the reported PWNe and PWNe candidates by \citet{HESS18_PWN, HESS18_HGPS}, the spatial extensions of some sources are significantly larger than $10\,$pc which are beyond the prediction of the dynamical evolution model for PWNe \citep{Reynolds84, vanderSwaluw01, Bucciantini03}. Some of these sources could also be pulsar halos in nature, or a mixture of PWN and halo \citep[e.g.][]{Liu20}. The Large High Altitude Air Shower Observatory (LHAASO) is a new generation instrument to study the TeV -- PeV \gray sky \citep{LHAASO19}. LHAASO contains two major \gray astronomic devices: the Water Cherenkov Detector Array (WCDA) and 1 $\rm km^2$ Array (KM2A). Recently, LHAASO reported discovery of 12 \gray sources above 100\,TeV, and some of them are pulsar halo candidates \citep{LHAASO21}. With its great sensitivity and large-sky area monitoring capability, LHAASO will play a key role in the detection of pulsar halos at TeV -- PeV band.

The morphology of the \gray halos is an important property for identifying pulsar halos. A pulsar's proper motion may lead to a displacement of about $v_pt_{\rm age}=80(v_{\rm p}/400{\,\rm km~s^{-1}})(t_{\rm age}/200{\,\rm kyr})$\,pc from its birth place. Such a displacement could be larger than the spatial extension of the TeV halo for middle-aged pulsars, so it may largely affect the \gray morphology. 
For example, Geminga has a transverse velocity of about $211(d/250\,\rm pc)$\,km/s \citep{Faherty07} and has travelled about 70\,pc in the plane of sky. Some research groups have discussed about the influence of Geminga proper motion to the morphology of its \gray halo \citep{Tang19, DiMauro19, Johan19}. It is found that the proper motion of Geminga can induce considerable asymmetry to its pulsar halo around 10\,GeV \citep{DiMauro19}, while the deviation of TeV halo from symmetry is of the order of 5\%-10\% only \citep{Johan19}. 
At the same time, spatial offsets between the centroids of candidates of pulsar halos/PWNe and the associated pulsars are commonly observed \citep{HAWC18, HAWC20_3HWC, HESS18_PWN}. Although the offset may be explained with the proper motion of the pulsar \citep{HESS18_PWN, DiMauro20}, it is not straightforward to connect a \gray source and the associated pulsar in the case of large offset being detected, especially without a knowledge on the pulsar's proper velocity. In addition, various factors such as continuous injection of electrons, diffusion and cooling of electrons,  magnetic and radiation field and the limit of the angular resolution of instruments (i.e., the point spread function, PSF) could influence the expected offset. This could complicate the identification of pulsar halos, while a detailed evaluation on the influence of these factors has not been performed.

Motivated by this, we study the effect of pulsar's proper motion on the  morphology of pulsar halos at GeV and TeV band. We focus on pulsar with an age of $t_{\rm age}\geq 100\,$kyr, so that the pulsar has escaped out of the SNR and we can assume electrons that escape the PWN are diffusing in the ISM \citep{Giacinti20}. For comparison with observations, we simulate the measured morphology of the halo by convolving the model-predicted intrinsic morphology with the PSF of some instruments, such as Fermi-LAT, HESS and LHAASO. The rest of the paper is organized as follows. In Section 2, we introduce our model for the \gray emission of the pulsar halo. In Section 3, we show and analyze the resulting \gray intensity map (morphology) of pulsar halo. In Section 4 we look into the offset between the center of the pulsar halo and the pulsar's location caused by the proper motion . We discuss the result in Section 5 and give our conclusion in Section 6.

\section{Model for the $\gamma$-ray emission of a pulsar halo}
In the model of pulsar halo considered in this work, relativistic electrons are continuously injected into the ambient ISM by a pulsar. We assume that the evolution of the total electron injection rate follows the pulsar's spin-down history with a braking index $n$ \citep{GS06}
\begin{equation}\label{eq:spin-down}
L_e(t)=\eta_e L_{i}{ (1+t/\tau_0)^{-(n+1)/(n-1)}},
\end{equation}
where $t$ is the time after the pulsar's birth, $\eta_e$ is the fraction of the spin-down energy converted into relativistic electrons, $L_i$ is the initial spin-down luminosity of the pulsar, $\tau_0=2\tau_c/(n-1)-t_{\rm age}$ is the initial spin-down timescale with $\tau_c$ being the ``characteristic age''. Given the present rotation period $P$ of the pulsar, the first derivative of the period $\dot{P}$ and the initial rotation period $P_0$, we have $\tau_c=P/2\dot{P}$ and $\tau_{\rm age}=\left[2\tau_c/(n-1)\right]\left[1-(P_0/P)^{n-1}\right]$.

The observed and derived values of above pulsar parameters scan a large range, e.g. $1<n<3$, their influence will be discussed in Section.~\ref{sec:separation} and Section.~\ref{sec:discussion}. For clarity, we define two benchmark pulsars for which the main results are calculated, unless differently specified. We assume the dipole magnetic model and take $n=3$. We refer to the value of $P$, $\dot{P}$ of Geminga and assume $P_0=50\,$ms, then we get $\tau_0=15\,\rm kyr$. We take $L_i=1.7\times10^{37}\,$erg/s, and notice that as a normalization, $\eta_e$ or $L_i$ will not influence the morphology of the pulsar halo. As for the age of the benchmark pulsar, we consider two cases, one is $t_{\rm age}=327\,\rm kyr$, corresponding to the age of Geminga, and the other is $t_{\rm age}=100\,\rm kyr$, for which the most energetic electrons have not cooled to several TeV.


\begin{table}[ht]
\centering
\caption{Properties of the benchmark pulsar.}\label{benchmark}
\setlength{\tabcolsep}{30pt}{
\begin{tabular}{lc}
\hline\hline
Parameter&Value\\
\hline
$L_i$\,/\,$\rm erg s^{-1}$&$1.7\times10^{37}\,\rm$\\
$\tau_0$\,/\,kyr&15\\
n&3\\
$t_{\rm age}$\,/\,kyr&100,327\\
$v_{\rm p}$\,/\,$\rm km s^{-1}$&400\\
d\,/\,kpc&2\\
\hline
\end{tabular}}
\tablecomments{The initial spin-down luminosity ($L_i$), initial spin-down timescale ($\tau_0$), braking index ($n$), age ($t_{\rm age}$), proper velocity ($v_{\rm p}$) and distance (d) of the benchmark pulsar that defined to calculate the morphology of pulsar halo.}
\end{table}

We assume the injection electron spectrum to be a power-law function with an exponential cutoff, i.e.,
\begin{equation}
Q_{e}(E_e,t)=Q_0(t)E_e^{-\gamma_e}\exp\left(-\frac{E_e}{E_c}\right),
\end{equation}
where $Q_0(t)$ is the normalization factor which can be determined by $\int E_e Q_e(E_e,t) dE_e=L_e(t)$. We consider  $\eta_e=1$, $\gamma_e=2$ and $E_c=400\,\rm TeV$ as reference parameters. For simplicity, we here do not consider any possible influence of the PWN and the related SNR on the historically injected electrons throughout the calculation. This might affect the resulting radiation at GeV band but is not supposed to influence the halo at  multi-TeV energy or beyond, since the emitting electrons of the latter cool within hundreds years and hence we do not expect to see the contribution of electrons injected at early epoch. The shape of assumed injection electron spectrum may have influence to the morphology of pulsar halo, which is discussed in Section.~\ref{sec:separation}. 

The transport equation of electrons injected from a point source located at $\pmb{r}_s$ is
\begin{equation}
    \frac{\partial n_e}{\partial t}=D(E_e)\nabla _{\pmb{r}}^2 n_e + \frac{\partial [b(\pmb{r},E_e,t)n_e]}{\partial E_e} +Q_e(E_e,t)\delta^3(\pmb{r}-\pmb{r}_s),
\end{equation}
where $n_e(\pmb{r}, E_e, t)$ is the differential electron density at time $t$ and position $\pmb{r}$, and $D(E_e)$ is the diffusion coefficient assuming isotropic diffusion, which is assumed to be spatially homogeneous and scales with the energy as $D(E_e)=D_0(E_e/1\rm GeV)^{1/3}$ following a Kolmogorov-type turbulence. We note that, based on HAWC's observation of Geminga and PSR B0656+14, it has been suggested that there may be a slow diffusion zone that $D_0\approx 10^{26}\,\rm cm^2/s$ in the vicinity of the pulsar within a radius of $30-100\,$pc from the pulsar while the diffusion coefficient beyond the radius is the standard one in the ISM measured from the primary-to-secondary cosmic ray ratio \citep{Fang18, Profumo18}. In this work, we do not consider the two-zone scenario but will only discuss its influence on our result qualitatively. 

The term $b(\pmb{r},E_e,t)$ is the energy loss rate of electrons during the  propagation due to the synchrotron  and IC radiation. We here consider a homogeneous and constant magnetic field and radiation field. The energy loss rate, $b$, is given by
\begin{equation}\label{eq:be}
b(E_e)=-\frac{d E_e}{d t}\simeq -\frac{4}{3}\sigma_Tc\left(\frac{E_e}{m_ec^2}\right)^2\left[U_B+\frac{U_{ph}}{(1+\frac{4E_e \epsilon_0}{m_e^2c^4})^{3/2}}\right],
\end{equation}
where $\sigma_T$ is the Thomson cross section, $U_B=B^2/8\pi$ is the magnetic field energy density and $U_{\rm ph}$ is the radiation field energy density. $\epsilon_0=2.82kT$ is the typical photon energy of blackbody/greybody target radiation field. We assume a homogeneous magnetic field of $B=3\,\rm\mu G$ and consider four blackbody/greybody components around the pulsar as: CMB ($T=2.73\,\rm K$ and $U=0.25\,\rm eVcm^{-3}$), far-infrared radiation (FIR) field ($T=40\,\rm K$ and $U=1\,\rm eVcm^{-3}$), near-infrared radiation field ($T=500\,\rm K$ and $U=0.4\,\rm eVcm^{-3}$) and visible light radiation field ($T=3500\,\rm K$ and $U=1.9\,\rm eVcm^{-3}$). The employed interstellar radiation field refers to the model by \citet{Popescu17} at a smaller Galactocentric radius (i.e., $3-5\,$kpc) given more TeV PWNe and PWN candidates appearing at such radius \citep{HESS18_PWN}. The influence of magnetic field and radiation field intensities to the morphology of pulsar halo will be discussed in Section.~\ref{sec:separation} and Appendix A. 

The present-day ($t=t_{\rm age}$) density of electrons with energy $E_e$ at a radius $r$ away from the pulsar can be calculated by 
\begin{equation}
\begin{split}
   n_{e} &(\pmb{r}, E_e) =\int_{0}^{t_{\rm age}} dt'Q_e(\mathcal{E}'_e(t'),t') \\
  &\times \frac{\exp[-(\pmb{r}-\pmb{r}_s(t'))^2/4\lambda(E_e,t')]}{(4\pi\lambda(E_e,t'))^{3/2}}
    \frac{d\mathcal{E}'_e(t')}{dE_e},
\end{split}
\end{equation}
where $\lambda(E_e,t') =\int_{t'}^{t_{age}}D(E_{e}(t''))dt''$ and $\mathcal{E}'_e$ is the electron energy at injection.  $E_e(t')$ represents the trajectory of energy evolution of an electron whose energy is $E_e$ at present. The relation between $E_e(t')$ and $\mathcal{E}'_e$ as well as $d\mathcal{E}'_e/dE_e$ can be found by tracing the energy evolution of the electron via Eq.~(\ref{eq:be}). 
We take the cylindrical coordinate system and set the direction of the proper motion as the $x-$axis. The present pulsar position is set to (0,0,0) and then the historical position of the pulsar can be give by $\pmb{r}_s=(v_{\rm p}(t_{\rm age}-t),0,0)$. The measurement of pulsar velocity along the line of sight (LOS) is generally not possible, except several binaries \citep{Hobbs05}. The mean transverse velocity of pulsars with $\tau_{\rm c}<10^3\,\rm kyr$ is 307\,km/s if modelling the pulsar velocity distribution with a Maxwellian distribution \citep{Hobbs05}. By modelling the velocity distribution with two Maxwellians, \cite{Verbunt17} found that 32 percent of the pulsars with $\tau_{\rm c}<10^4$\,kyr have an average transverse velocity of $130 {\,\rm km/s}$ and 68 percent with $520\,\rm km/s$. For our benchmark pulsar, we take the angle between pulsar proper motion direction and LOS as $\psi=90^\circ$ and take $v_{\rm p}=400\,\rm km/s$, namely the transverse velocity of pulsar $v_{\rm tr}$, to explore the morphology of pulsar halo at different energies. We will discuss the influence of the value and direction of the pulsar proper velocity to the morphology of pulsar halo in Section.~\ref{sec:separation}.

We then calculate the \gray emission produced by the IC process of electrons and integrate them over the LOS at different viewing angle, following the method given by \citet{Liu19_prl}. We assume the distance of benchmark pulsar to be 2\,kpc, and will discuss its influence in Section.~\ref{sec:separation}.
By taking $\psi=90^\circ$, the emission of the halo is projected onto the celestial plane and we obtain the \gray intensity at a polar angle $\theta$ from the pulsar and an azimuth angle $\phi$ ($\phi=0$ points to the opposite direction of the proper motion) $I_\gamma'(E_\gamma,\theta, \phi)$, which describes the intrinsic morphology of the \gray halo.

We convolve $I_\gamma'$ with the PSF of various instruments at different energies (summarized in Table.~\ref{psflist}) to simulate the observed intensity map or the morphology of the \gray halo by the instruments. The \gray intensity map after the convolution with the PSF is then given by
\begin{equation}
I_{\gamma}(E_\gamma,\theta,\phi)=\iint\frac{1}{2\pi \sigma^2} \exp\left(-\frac{l'^2}{2\sigma^2}\right) I_\gamma'(E_\gamma,\theta',\phi') \sin\theta'd\theta'd\phi'.
\end{equation}
where $l'=\cos\theta \cos\theta'+\sin\theta \sin\theta' \cos(\phi-\phi')$ is the angular distance between the point $(\theta,\phi)$ and the point $(\theta',\phi')$ in the celestial plane, and $\sigma(E_\gamma)$ is the size of the PSF as a function of \gray energy. 

\begin{figure*}[htb]
\centering
\includegraphics[width=1.0\textwidth]{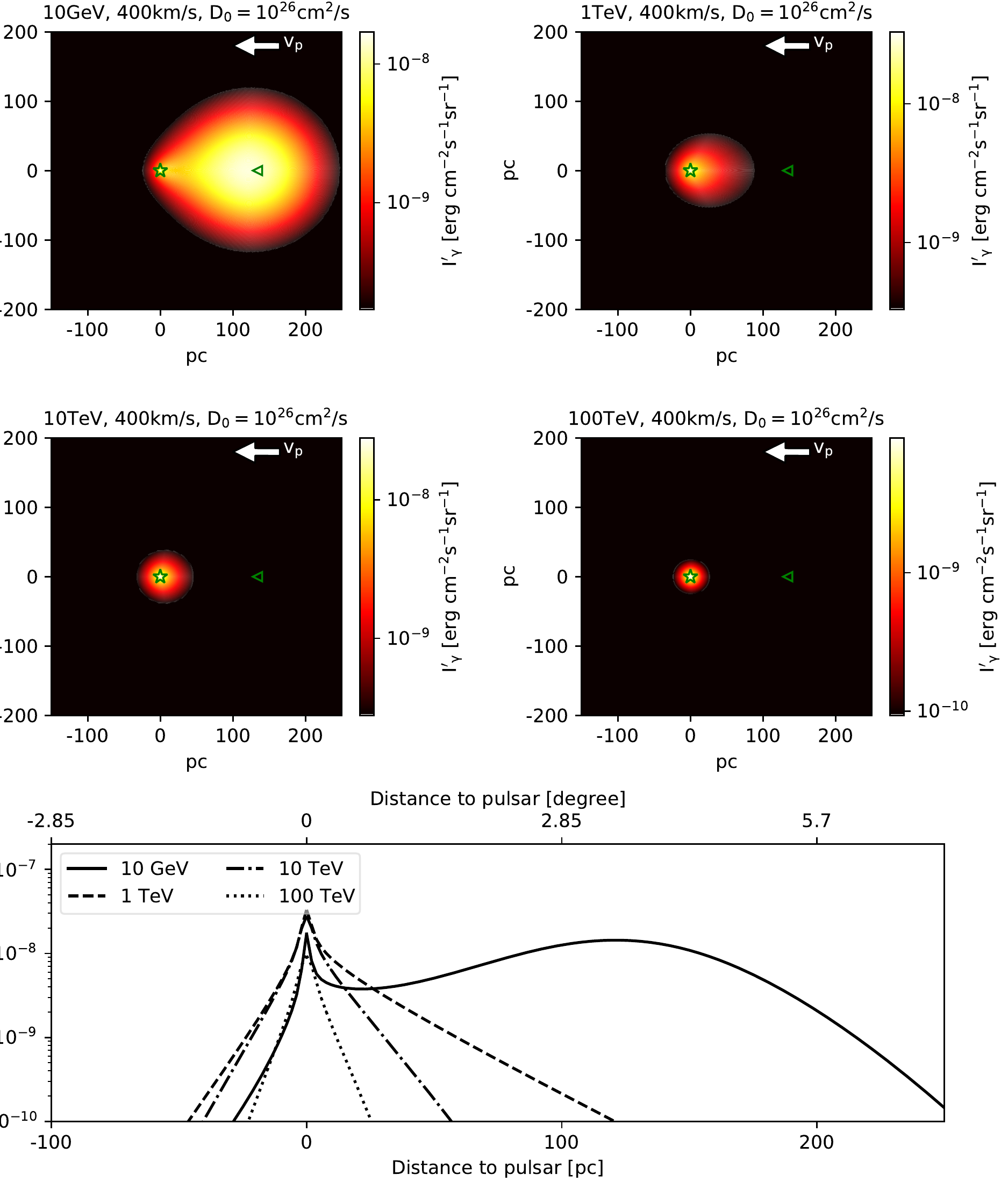}
\caption{\gray intensity map of pulsar halo and intensity profile along the axis of pulsar's proper motion at $10\,\rm GeV$ and $1,10,100\,\rm TeV$, considering the proper motion of the benchmark pulsar ($t_{\rm age}=$327\,kyr) at 2\,kpc. The direction of proper motion is marked with the arrow. The star and triangular mark the present and initial position of pulsar. The proper velocity is $v_{\rm p}=400\,\rm km/s$, $\psi=90^\circ$ and the diffusion coefficient is $D_0(1\,\rm GeV)=10^{26}\,\rm cm^2/s$.}
\label{morphology}
\end{figure*}

\section{Morphology of the Pulsar halo}\label{sec:morphology}

We first simulate and compare the morphology of the \gray halo of the benchmark pulsar ($t_{\rm age}=$327\,kyr) with the results of \citet{Tang19} and \citet{DiMauro19}, which considered continuous injection of electrons and one-zone diffusion model. Our result is consistent with theirs qualitatively, though we assume different properties of the benchmark pulsar. Specifically, the intrinsic intensity maps of the pulsar halo at 10\,GeV, 1\,TeV, 10\,TeV and 100\,TeV, as well as the corresponding intensity profile along the axis of the pulsar's proper motion ($x$-axis), are shown in Fig.\ref{morphology}. The influence of pulsar proper motion to the morphology of its \gray halo is energy-dependent.
The morphology of the pulsar halo at 10 GeV and 1 TeV is highly asymmetrical, with extended emission towards the right half part where the pulsar has passed. At 10\,GeV, the pulsar halo is double-peak. The strongest emission is around the pulsar's initial position, where the relic electrons injected at the early epoch are still emitting $\gamma$-rays. This is because the cooling time of GeV-emitting electrons is longer than the age of the pulsar and the spin-down luminosity of the pulsar at the early epoch is hundreds times larger than that at present. On the other hand, electrons injected recently form a comparatively high electron density region around the pulsar since they have not diffused far away, thus yielding another peak in the intensity profile.
For TeV-emitting electrons, their cooling time is shorter than, but still comparable to, the age of the pulsar. Therefore, the intensity map of the pulsar halo at 1\,TeV extends towards $x>0$ but peaks around the current position of the pulsar. At higher-energy (e.g., 10\,TeV and 100\,TeV), the relic electrons are significantly cooled so the morphology of the halo shows a single peak centered at the current position of the pulsar without a significant extension and the emission of halo is dominated by recently injected electrons.

\begin{figure*}[htb]
\centering
\includegraphics[width=1.0\textwidth]{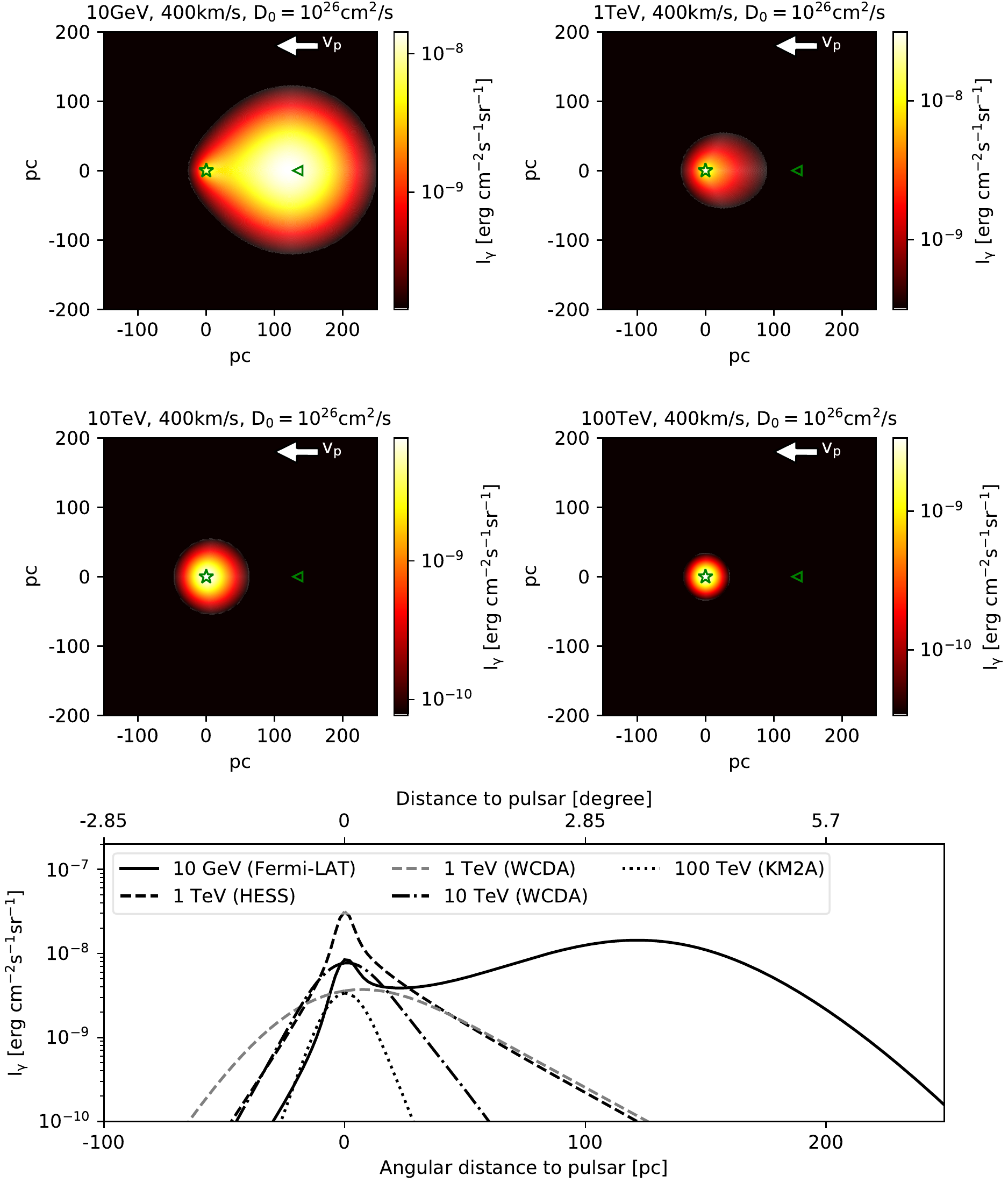}
\caption{Same as Fig.~\ref{morphology} but after convolving with the PSF of instruments. We take the PSF of Fermi-LAT (10\,GeV), HESS (1\,TeV), WCDA (1,10\,TeV) and KM2A (100\,TeV), respectively.}
\label{morphology_psf}
\end{figure*}

\begin{table}[b]
\centering
\caption{PSF of \gray detectors at different energies.}\label{psflist}
\begin{tabular}{c|l|ccccc}
\hline\hline
\multirow{2}{*}{Fermi-LAT} & $E_\gamma$\,[GeV] & 1.6 & 2.6 & 5.2 & 10.4 & 52 \\
~ & PSF\,[$^\circ$] & 0.4 & 0.26 & 0.16 & 0.1 & 0.06 \\
\hline
\multirow{2}{*}{HESS} & $E_\gamma$\,[TeV] & 1.0 & 3.3 & 6.6 & 8.3 & 16.5 \\
~ & PSF\,[$^\circ$] & 0.07 & 0.06 & 0.05 & 0.05 & 0.04\\
\hline
\multirow{2}{*}{WCDA} & $E_\gamma$\,[TeV] & 1.0 & 3.3 & 6.6 & 8.3 & 16.5 \\
~ & PSF\,[$^\circ$] & 0.6 & 0.34 & 0.28 & 0.27 & 0.24\\
\hline
\multirow{2}{*}{KM2A} & $E_\gamma$\,[TeV] & 16 & 26 & 39 & 59 & 88 \\
~ & PSF\,[$^\circ$] & 0.5 & 0.37 & 0.3 & 0.24 & 0.2\\
\hline
\end{tabular}
\end{table}

In order to compare with observations, we also give the intensity map and the profile after the convolution with PSFs of four $\gamma$-ray dectors in Fig.~\ref{morphology_psf}, i.e., the PSF of Fermi-LAT for 10\,GeV emission, HESS for 1\,TeV emission, LHAASO(WCDA) for 1\,TeV and 10\,TeV emission and LHAASO(KM2A) for 100\,TeV emission (see Table.~\ref{psflist}). The PSF of HAWC is similar to that of LHAASO(WCDA). After the convolution, the profile becomes more smooth and extended. Such a change is more pronounced for larger PSF as can be seen by comparing the 1\,TeV profile convolved with WCDA's PSF and that convolved with HESS' PSF. 

To understand more quantitatively the energy-dependent morphology of the pulsar halo and the influences of different model parameters, let us first define two critical timescales. The first one is $t_{\rm pd}=80(E_e/1\,{\rm TeV})^{1/3}(D_0/{10^{26}\rm cm^2s^{-1}}) (v_{\rm tr}/400\,\rm km/s)^{-2}\,\rm kyr$, the time when the electron's diffusion distance ($2\sqrt{Dt}$) is equal to the distance that the pulsar has moved ($v_{\rm tr}t$). The other one, denoted by $t_{\rm c}$, is the time in which an electron cools from energy $E_c$ (the cutoff energy in the injection spectrum) to certain energy $E_e$ in the considered magnetic field and the radiation field. $t_{\rm c}$ can be calculated via Eq.~(\ref{eq:be}). For a fixed magnetic field and radiation field, the value of $t_{\rm c}$ is larger than the standard definition of the cooling timescale $\tilde{t}_{\rm c}(E_e)$ of electrons (i.e., $\tilde{t}_{\rm c}\equiv E_e/b(E_e)$). Its value is mainly determined by $E_e$ instead of $E_c$. Then we can divide the morphological evolution of the pulsar halo into three phases, based on these two timescales. This definition of phases can help to generalize different feature and origin of different pulsar halo morphology. This simple physical definition works well for pulsars with $n>2.5$ and $\tau_0>5\,$kyr. The three phases of pulsar halo evolution can be given by:

{\bf PHASE \uppercase\expandafter{\romannumeral 1}}: $t_{\rm age}< t_{\rm pd}, t_{\rm age}< t_{\rm c}$. The electrons injected at early epoch with a high luminosity haven't cooled and pulsar's displacement due to proper motion is still within the diffusion length of these relic electrons. The relic electrons produce bright radiation around the original position of the pulsar. Since relic electrons have diffused to a larger distance than the pulsar has travelled, the emissions of electrons injected at different epoch overlap with each other and yield a single and broad peak in the intensity map. 

{\bf PHASE \uppercase\expandafter{\romannumeral 2}}: $t_{\rm pd}< t_{\rm age}< t_{\rm c}$. Since the displacement of the pulsar from its original position is proportional to $t$ while electron's diffusion distance is proportional to $t^{1/2}$, the pulsar eventually goes beyond the diffusion length of electrons injected at early epoch, and the fresh electrons yield elongated bright regions along the pulsar trajectory. The pulsar halo is highly asymmetrical, being double peaked or single peaked with an elongated tail towards the direction $x>0$, such as the 10\,GeV intensity maps in Fig.\ref{morphology_psf}. The relative brightness at the original position and the current position of the pulsar depends on $L_e(t)$.

{\bf PHASE \uppercase\expandafter{\romannumeral 3}}: $t_{\rm age}> t_{\rm c}$. Electrons injected at early epoch lost most energy and only recently injected electrons can produce \gray emission. The morphology of pulsar halo is again single-peak and compact, with a rough circular symmetry with respect to the current position of the pulsar, such as the 10\,TeV and 100\,TeV intensity maps in Fig.\ref{morphology_psf}. 

\begin{figure}
\centering
\includegraphics[width=1.0\columnwidth]{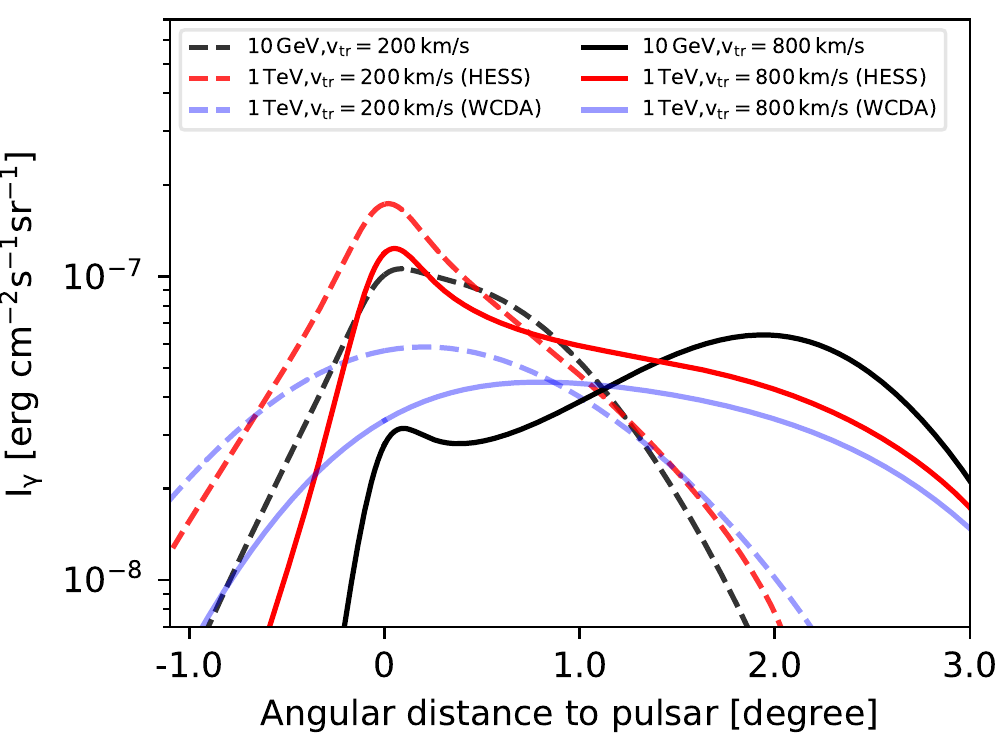}
\caption{The \gray intensity profile of benchmark pulsar ($t_{\rm age}=$100\,kyr), but with $v_{\rm tr}=200\,$km/s and $800\,\rm km/s$, after convolving with the PSF of Fermi-LAT (10\,GeV), HESS (1\,TeV) and LHAASO(WCDA) (1\,TeV).}
\label{profileyr}
\end{figure}

Note that in the transition stage from one phase to another phase (i.e., $t_{\rm age} \sim t_{\rm pd}$, $t_{\rm pd}\sim t_{\rm c}$), the feature of a certain phase mentioned above is not distinct and relies on the relative contribution from relic and newly injected electrons, i.e. $L_e(t)$. The halo of benchmark pulsar ($t_{\rm age}=$327\,kyr) is already in PHASE II at GeV band and PHASE III at TeV band. We show in Fig.\ref{profileyr} the intensity profiles of the pulsar halo of the benchmark pulsar ($t_{\rm age}=$100\,kyr) and assume $v_{\rm tr}=200\,$km/s and 800\,km/s respectively, to compare the feature of pulsar halo at PHASE I with that at PHASE II.
At this age, the pulsar halo at 10\,GeV and 1\,TeV with $v_{\rm tr}=200\,$km/s belongs to PHASE \uppercase\expandafter{\romannumeral 1} (dashed curves) while those with $v_{\rm tr}=800\,$km/s belong to PHASE \uppercase\expandafter{\romannumeral 2} (curve lines). As we discussed above, the pulsar halo at PHASE I is single-peak. The pulsar halo at PHASE II is more asymmetrical with an extension towards the direction of the pulsar's initial position.

\section{Separation angle estimation}\label{sec:separation}
We now look into the expected offset of the pulsar and the \gray halo at different energies when measured by different instruments. We first produce the PSF-convolved intensity maps at different energies by convolving $1-100$\,GeV emission with the PSF of Fermi-LAT, $1-20\,$TeV emission with the PSFs of LHAASO(WCDA) and HESS respectively, and $10-100\,$TeV emission with the PSF of KM2A. Then, we define two kinds of offsets: one is the separation between the pulsar' current position and the centroid of the halo (denoted by $\Theta$), and the other is the separation between the pulsar and the position of the brightest point or the intensity peak of the halo (denoted by $\Theta'$). The latter is straightforward to obtain. To get the former, we use a 2D Gaussian template as $I_G(\pmb{r})=(N_0/2\pi \sigma_0^2) \exp[-(\pmb{r}-\pmb{r}_c)^2/2\sigma_0^2]$, where $\pmb{r}_c$ is its center, to model the PSF-convolved intensity map. For each fitting, we adjust the size of region of interest, to include the whole pulsar halo (defined as the \gray intensity at the edge is 1\% of the brightest point) and at the same time, maintain a higher resolution. We set $N_0$, $\sigma_0$ and $\pmb{r}_c$ as free parameters, and search for the centroid of the halo ($\pmb{r}_c$) by minimizing the chi-square $\chi^2=\sum_{\pmb{r}}(I_G-I_\gamma)^2/I_\gamma$. This procedure can mimic the real analysis of observation, except there is no background emission in our simulation.

\begin{figure}[htbp]
\centering
\includegraphics[width=1.0\columnwidth]{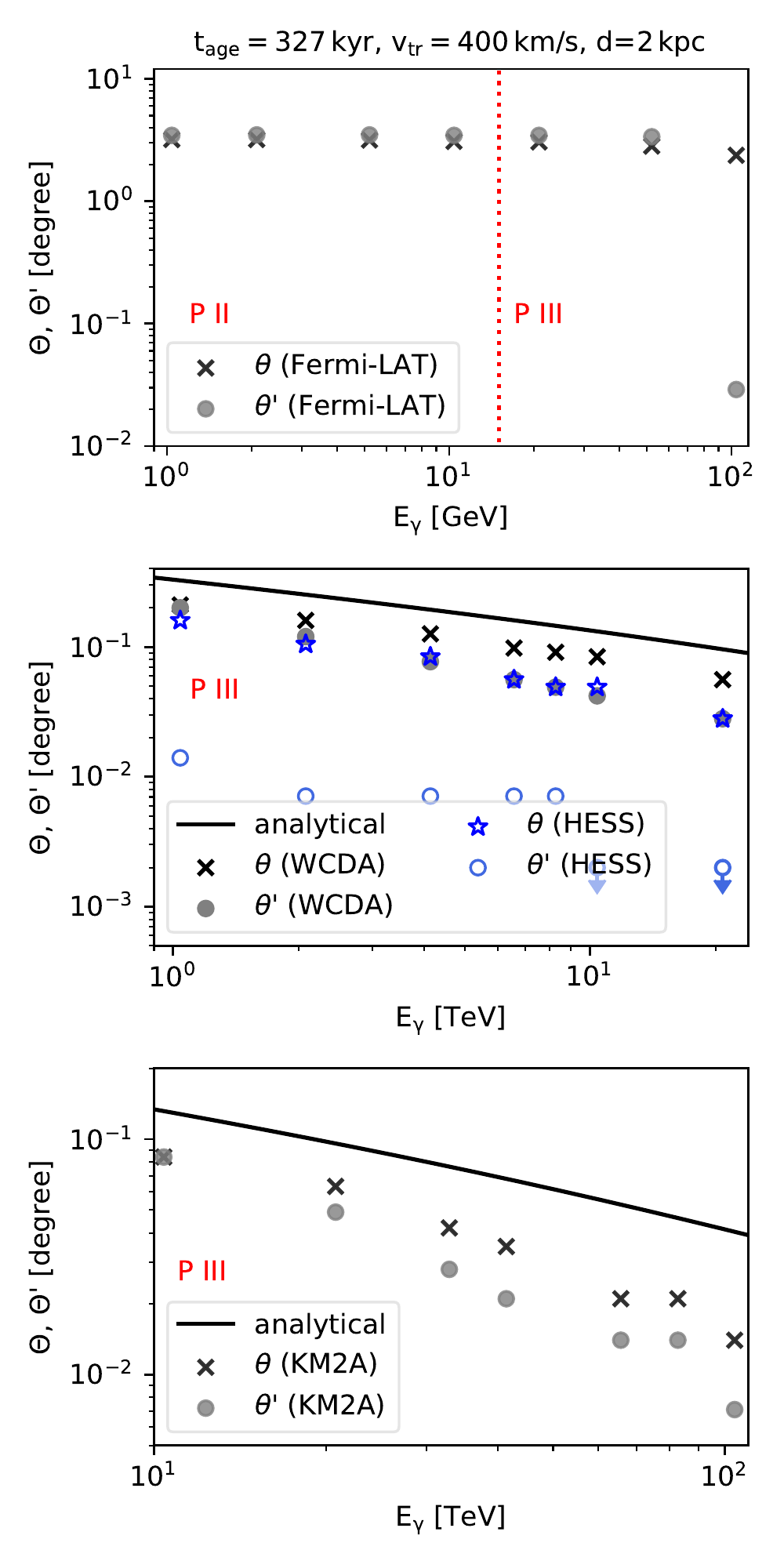}
\caption{The separation angle $\Theta$ defined by modelling Gaussion template (crosses and stars) and $\Theta'$ defined as peak of halo (dots and circles) at different energies of benchmark pulsar ($t_{\rm age}=$327\,kyr). We take PSF of Fermi-LAT for GeV, for $1-20\,\rm TeV$ we adopt PSF of WCDA and HESS, and PSF of KM2A for $10-100\,\rm TeV$ separately. The black line shows the analytical approximation of $\Theta'$ \citep{DiMauro20}. We annotate evolutionary phase of pulsar halo in red (i.e., P\,II for PHASE II and P\,III for PHASE III).}
\label{thetaE}
\end{figure}

Convolving the same intrinsic intensity map with the PSF of HESS and the PSF of WCDA respectively (for example, comparing the solid red and the solid blue lines in Fig.~\ref{profileyr}), the peak of the obtained intensity profile can be different.
This is also reflected in the separation angle which can be seen in the middle panel of Fig.~\ref{thetaE}. Both $\Theta$ and $\Theta'$ estimated with the PSF of HESS (blue markers) are smaller than those with the PSF of WCDA (black markers). Such a tendency is particularly pronounced if the offset is evaluated by $\Theta'$. For halo beyond TeV energy, the position of the intensity peak in the intrinsic intensity map is basically the position of the pulsar's current position. However, due to the asymmetry of the halo, the peak would shift to the more elongated side of the halo after convolving with the PSF of the detector. The smaller the PSF is, the less distance the peak would shift. For detector of good angular resolution, such as HESS, the induced $\Theta'$ could be even smaller than the resolution of our simulation if the intrinsic morphology of the halo is not sufficiently asymmetric.

The intriguing dependence of the resulting $\Theta$ and $\Theta'$ on various parameters, in addition to the PSF, will be discussed in the following subsections. When discussing the influence of parameters like diffusion coefficient, magnetic field and other parameters, we prefer to take the benchmark pulsar ($t_{\rm age=}100\,kyr$), thus the high energy electrons producing TeV \gray emission have not cooled and the influence of parameters can show up.
 
\subsection{Dependence on the \gray energy}

The obtained separation angles $\Theta$ and $\Theta'$ at different energies for benchmark pulsar ($t_{\rm age}=$327\,kyr) are plotted in Fig.\ref{thetaE}. To better interpret the results, we mark the evolutionary phases in each panel. The pulsar halo phase evolves following \uppercase\expandafter{\romannumeral 2} $\rightarrow$ \uppercase\expandafter{\romannumeral 1} $\rightarrow$ \uppercase\expandafter{\romannumeral 3} or directly II $\rightarrow$ III as the energy increases from GeV to 10\,TeV. The separation angle here is calculated based on a nominal distance of $d=2$\,kpc for the pulsar, and the result approximately scales with $d^{-1}$.

At GeV band $\Theta$ is energy-independent and is close to the displacement of the pulsar due to proper motion in a time of  $t_{\rm age}$, which is $\Theta =3.8^\circ$ for a nominal distance of 2\,kpc of the pulsar. It reflects the emission of the huge amount of relic electrons injected at early epoch (PHASE \uppercase\expandafter{\romannumeral 2} and transition to PHASE III). $\Theta'$ is constant at PHASE II but jumps to almost zero after the pulsar halo entering PHASE III. The reason for such a behavior is as follows: As the pulsar halo transiting from PHASE II to PHASE III, the relic electrons start to cool so the radiation around the initial position of the pulsar becomes dimmer. At $\sim 100$\,GeV, the relic electrons after cooling still make an important contribution to the entire emission of the pulsar so the decrease of $\Theta$ is limited, but the peak intensity is already lower than that of the electrons freshly injected, so $\Theta'$ is nearly zero.
As energy further going up, the cooling of emitting electrons becomes increasingly important (PHASE \uppercase\expandafter{\romannumeral 3}), the contribution of relic electrons reduces, leading to a more symmetric morphology of the halo at higher energy. The separation, either defined by $\Theta$ or $\Theta'$, becomes smaller. Above 10\,TeV the separation is smaller than $0.1^\circ$ and is difficult to be resolved for a pulsar at 2\,kpc by current gamma-ray detectors, especially for those with large PSFs such as HAWC and LHAASO. 

\citet{DiMauro20} suggested that the offset between pulsar and pulsar halo induced by pulsar proper motion can be analytically depicted by $\Theta'(E_\gamma)={\rm atan}(v_{\rm p} \tilde{t}_{\rm c}(E_\gamma)/d)$ in the cooling-dominated stage (i.e., deep in PHASE III), where $\tilde{t}_{\rm c}(E_\gamma)$ is the cooling timescale of electrons that dominates the gamma-ray flux at $E_\gamma$. We overlaid the result given by this formula\footnote{The formula given by \citet{DiMauro20} is an approximation and does not take into account the Klein-Nishina effect, which is important at high energy. Note that the Klein-Nishina effect is considered in their numerical simulations.} in Fig.~\ref{thetaE} for reference. The formula generally reproduces the trend of $\Theta$ v.s. \gray energy, but it overestimates the separation angle. This is due to the ignorance of the influence of the continuous injection of electrons, which would produce \gray emission along pulsar trajectory and make the center of pulsar halo deviate away from the position defined in the analytical formula. The influence of the continuous injection will be further discussed in Section~\ref{sec:max}.

\subsection{Dependence on the distance, proper velocity and diffusion coefficient}

\begin{figure}[h]
\centering
\includegraphics[width=0.9\columnwidth]{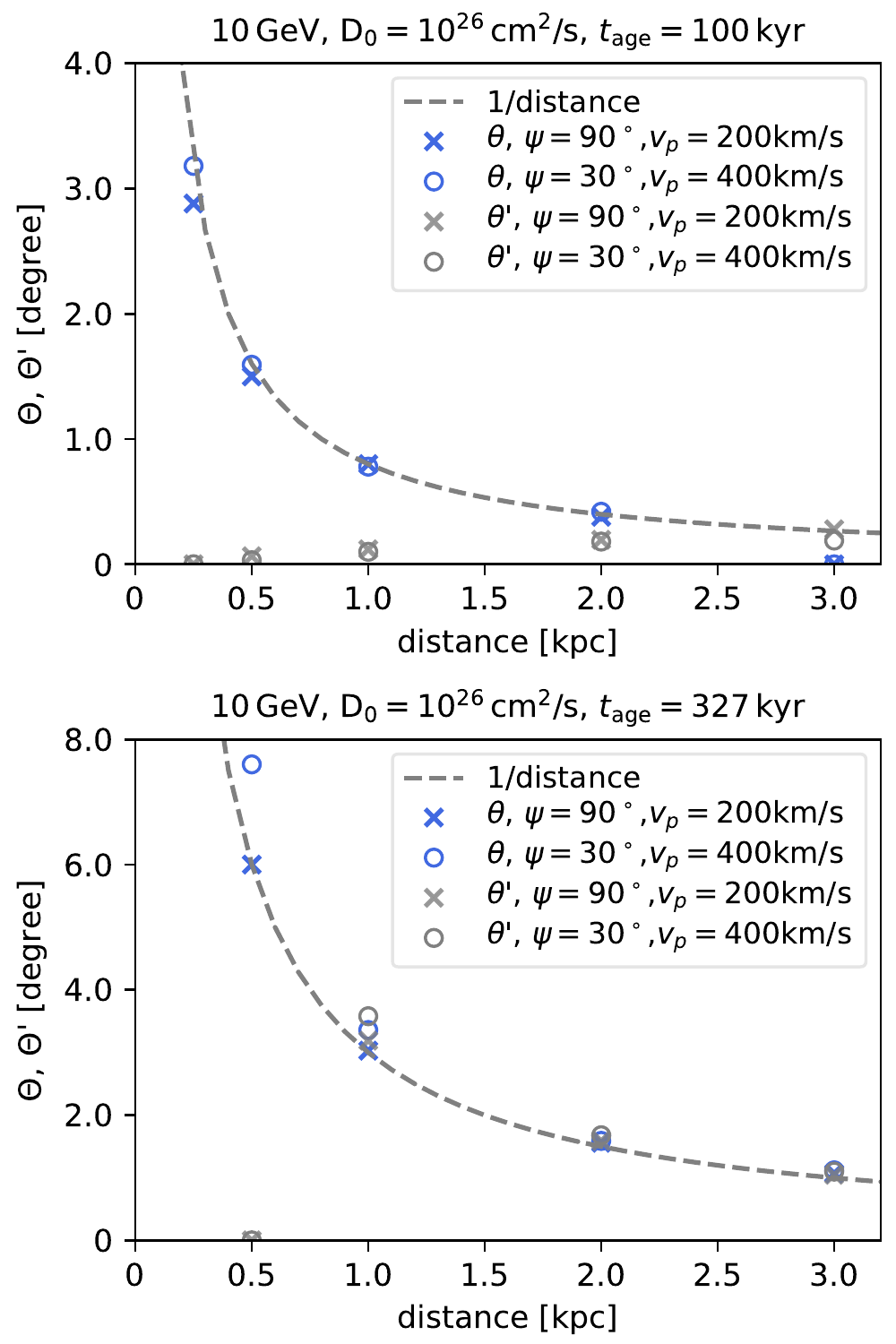}
\caption{The separation angle $\Theta$ and $\Theta'$ of the 10\,GeV halo as a function of pulsar's distances with the same transverse velocity  but different LOS velocity of the pulsar. Crosses represent the case with $v_p=200\,\rm km/s$ and $\psi=90^\circ$, while circles represent the case with $v_p=400\,\rm km/s$ and $\psi=30^\circ$. The upper and lower panels show the separation angles at $t_{\rm age}=100\,$kyr and $t_{\rm age}=327\,$kyr respectively.}
\label{fig:thetavlos}
\end{figure}

\begin{figure*}[htbp]
\centering
\includegraphics[width=0.9\textwidth]{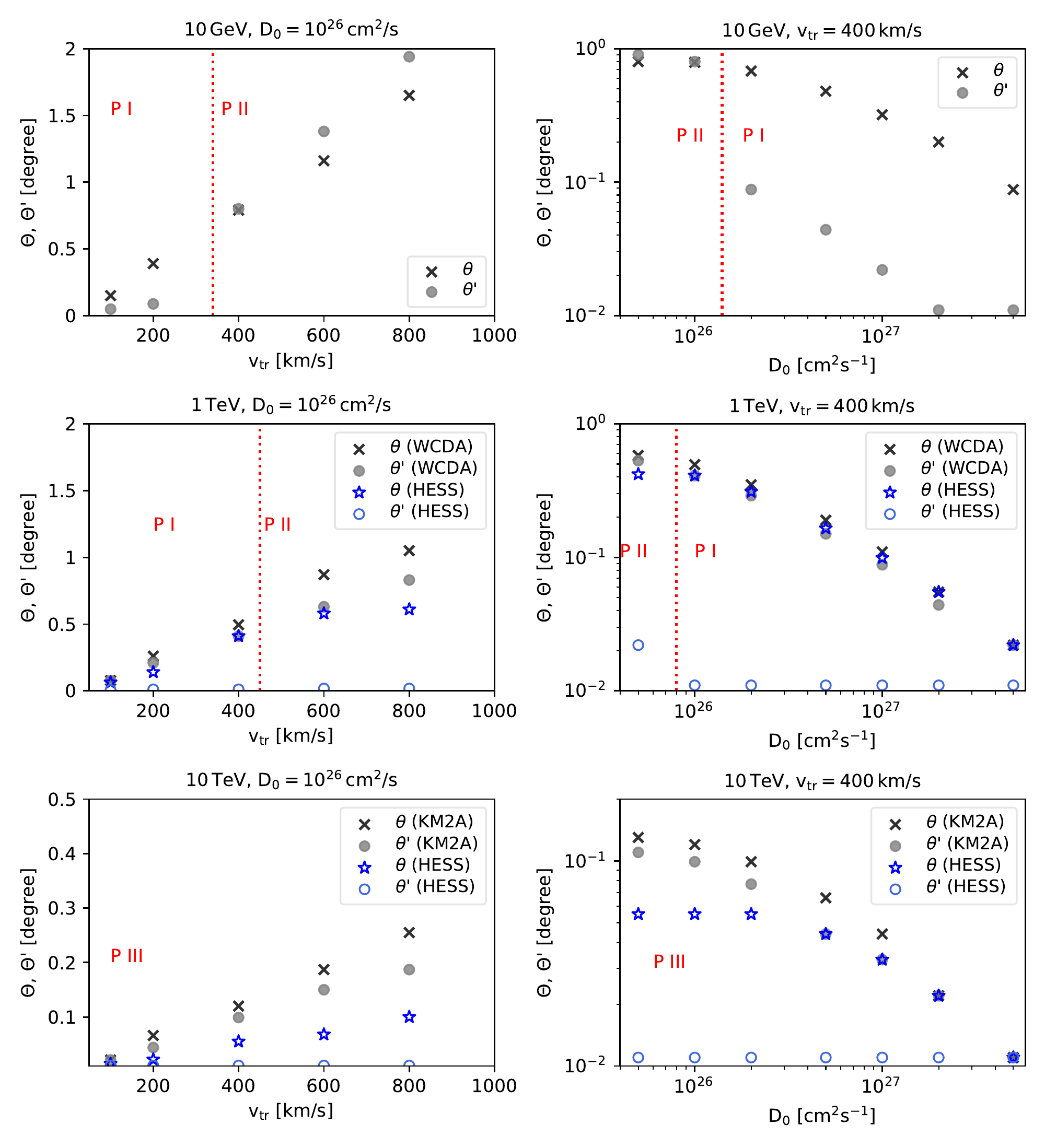}
\caption{The separation angle $\Theta$ and $\Theta'$ at different transverse velocity $v_{\rm tr}$ (left panels) and diffusion coefficient $D_0$ (right panels) for the benchmark pulsar ($t_{\rm age}=$100\,kyr) at 10\,GeV, 1\,TeV and 10\,TeV. The evolutionary phase of pulsar halo is annotated in red and red dotted lines show the boundary between two different phases.}
\label{fig:theta}
\end{figure*}

For the same pulsar, the separation angle of the pulsar halo is supposed to scale with the distance of the pulsar $\rm d_{kpc}$ as $\rm \Theta(d_{kpc})=\Theta(1\, kpc)/d_{kpc}$. In Fig.~\ref{fig:thetavlos}, we plot the calculated separation angle for the pulsar at different distances and the theoretical relation (dashed line). The relation fits the separation angles well except for $\Theta'$ of benchmark pulsar at $t_{\rm age}=$100\,kyr, because these $\Theta'$ are dominated by the convolution with the PSF of the instrument, when the intrinsic separation is very small. 

The direction of the proper velocity $v_{\rm p}$ of all pulsars in the galaxy is supposed to be randomly distributed, i.e. the cosine of the angle between $v_{\rm p}$ and LOS, $\psi$, is equally distributed in the range $(0,1)$. Considering the observed gamma-ray flux is proportional to $1/d^2$, the gamma-ray luminosity could be larger at the initial position if the pulsar has a radial velocity and moves away from us, i.e. $\psi<90^\circ$, vise versa. Therefore the LOS velocity, which is equal to $v_{\rm p}\cos\psi$ may affect the pulsar halo morphology and separation angle we observed. We consider two cases for the pulsar's proper motion: one is $v_p=200\,\rm km/s$ with $\psi=90^\circ$, and the other is $v_p=400\,\rm km/s$ with $\psi=30^\circ$, i.e. the transverse velocity in the plane of sky $v_{\rm tr}=v_p\sin\psi$ of the two cases are the same. The separation angles of pulsars with these two velocities at different distances are plotted in Fig.~\ref{fig:thetavlos}. We find that the direction of $v_{\rm p}$ only causes observable influence to the separation angle if the length that pulsar moved along LOS is comparable to the distance of pulsar. Thus for the pulsar at larger distance, we can ignore its velocity along LOS and simply take its observed $v_{\rm tr}$ to calculate the theoretical gamma-ray halo morphology. 

Since electrons are continuously injected into the ambient medium of the pulsar, a larger $v_{\rm tr}$ would induce a more elongated tail of the halo towards $x>0$, as Fig.~\ref{profileyr} shown. This is correspondingly reflected in the $\Theta(\Theta') - v_{\rm tr}$ relation, as a monotonous increase of the separation angle with the proper velocity\footnote{An exception is the $\Theta'$ for HESS as shown with open blue circles. The offset between the position of the peak and the position of the pulsar can be hardly identified in our simulation due to its small PSF. See also the discussion in the beginning of Section~\ref{sec:separation}.}. 
The transition of the slope of the relation can be explained by the transition of morphology of halo. For example, the jump of $\Theta'$ with increasing $v_{\rm tr}$ for the 10\,GeV halo (the top left panel of Fig.~\ref{fig:theta}) is due to that the pulsar halo transits from single peak to double peak (i.e., PHASE I to PHASE II). The reason is the same for the jump of $\Theta'$ with increasing $D_0$ for the 10\,GeV halo (top right panel of Fig.~\ref{fig:theta}). The other interesting transition occurs when the Gaussian template is used to estimate the offset and the intensity profile is convolved with HESS PSF, as shown with blue stars in middle left panel of Fig.~\ref{fig:theta}, the $\Theta-v_{\rm tr}$ relation becomes flat when $v_{\rm tr}>400\,\rm km/s$. The reason is that the morphology of the intrinsic halo at TeV band has a single broad peak in PHASE I, which can be described with a Gaussian template. In PHASE II, the morphology of the halo is single peaked with an elongated tail at TeV band. Due to the good resolution of HESS (i.e., small PSF), such a feature is reserved in the PSF-convolved intensity profile and deviates from the Gaussian template. Therefore, when using the Gaussian template to fit the intensity map measured by HESS, the peak is over-weighted while the elongated tail barely influences the determination of the centroid. By contrast, when the intensity profile in PHASE II is convolved with the PSF of WCDA, the latter's large PSF smears out the fine structure, leaving only a single broad peak (see the solid blue curve in Fig.~\ref{profileyr}) which is similar to  that in PHASE I.  As a result, we do not expect to see such a prominent transition in the $\Theta-v_{\rm tr}$ relation if the halo is measured by WCDA, as shown with the black crosses in the middle left panel of Fig.~\ref{fig:theta}. 

With decreasing diffusion coefficient, electrons injected at early epoch diffuse more slowly and yield a brighter region at $x>0$, so we expect a more asymmetric pulsar halo and subsequently a larger separation angle with a smaller $D_0$, as the right panels of Fig.~\ref{fig:theta} shown. Specifically, if the emitting electrons at the beginning stage are not cooled, a smaller $D_0$ results in a smaller $t_{\rm pd}$, leading to the earlier transition of the halo from PHASE I to PHASE II. The radiation from the relic electrons and the fresh electrons becomes distinct as electrons are better confined around the position where they are injected, so the separation angle becomes larger. Similarly, although no relic electrons are survived in PHASE III, a smaller $D_0$ can nevertheless enhance the radiation of electrons injected at earlier time so the halo would appear more elongated toward $x>0$. However, we note that the influence of $D_0$ on $\Theta$ or $\Theta'$ is limited by the age of the pulsar for PHASE II or by the cooling timescale of electrons in PHASE III, because the separation angle would be at most $v_{\rm tr}t_{\rm age}/d$ for PHASE I/II or $v_{\rm tr}t_{\rm c}/d$ for PHASE III. 
This explains why the slope of $\Theta(\Theta')$ against diffusion coefficient become flat at the low $D_0$ end in the right panels of Fig.~\ref{fig:theta}. 

\subsection{Dependence on the injection spectrum of electrons}

When the early injected high-energy electrons cool to lower energies, they can produce lower-energy \gray emission. Therefore, the electron injection spectrum may affect the separation angle of the pulsar halo.
Here we compare two additional cases of power-law electron injection spectrum, one softer ($\gamma_e=2.5$) and the other harder ($\gamma_e=1.5$), with that in the benchmark case ($\gamma_e=2.0$).  

\begin{figure*}[t]
\centering
\includegraphics[width=0.9\textwidth]{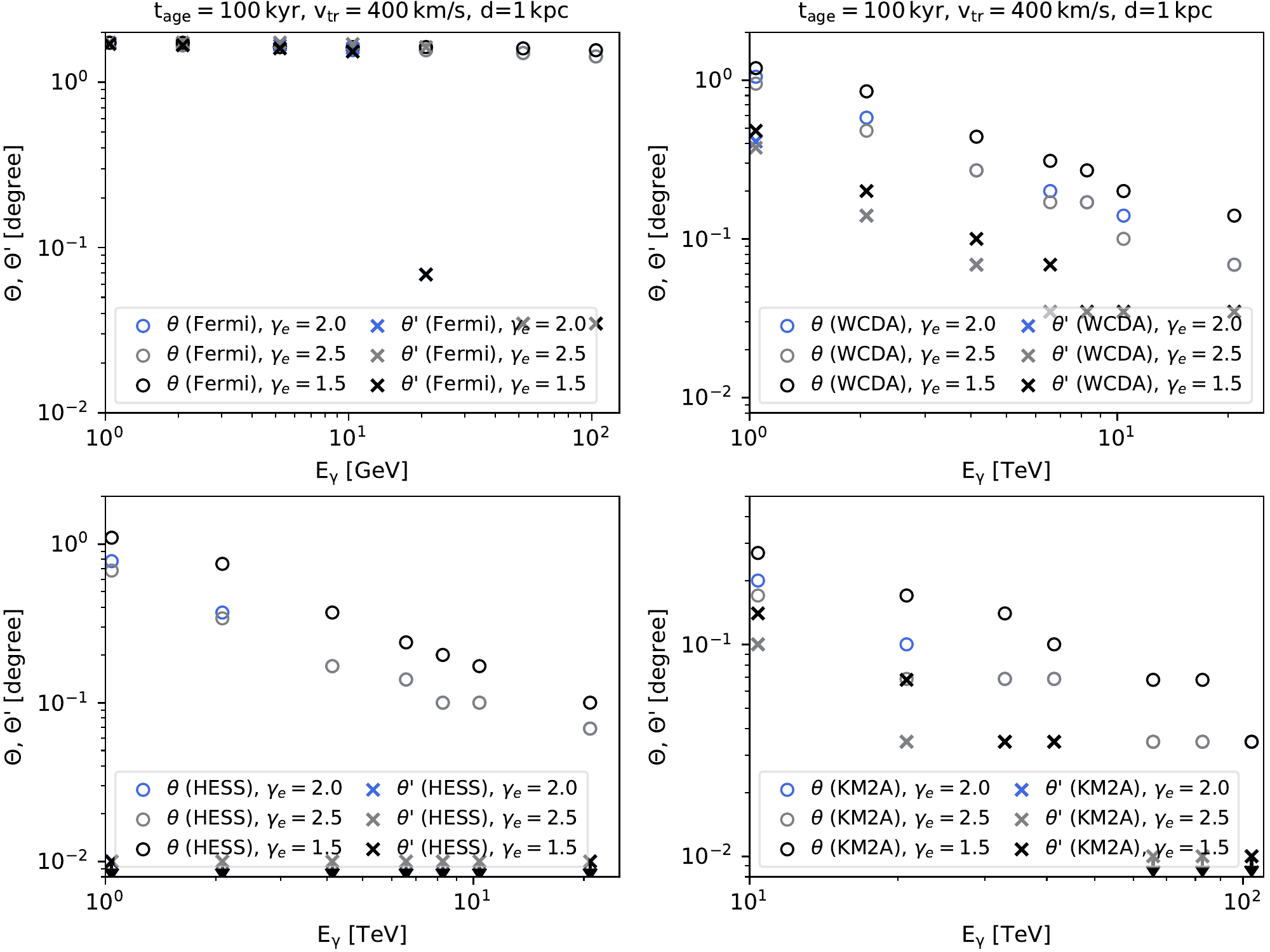}
\caption{The separation angle $\theta$ and $\theta'$ of the benchmark pulsar ($t_{\rm age}=$100\,kyr) with different electron injection spectrum at different energies. The blue markers are the results of power-law spectrum ($\gamma_e=2.0$) with an exponential cutoff. The gray markers are the results of softer power-law spectrum ($\gamma_e=2.5$). The black markers are the results of harder power-law spectrum ($\gamma_e=1.5$). Where no blue points is because blue points overlaid with gray points.}
\label{fig:thetaspe}
\end{figure*}

The gamma-ray emission at the initial position of pulsar is produced by electrons cooled from higher energy and a harder spectrum in general leads to more high-energy relic electrons. As such, the emission at the initial position is enhanced given a harder spectrum with respect to that at pulsar's current position. This is consistent with the results for $E>1\,$TeV as shown in Fig.~\ref{fig:thetaspe}, where we see the separation angles with $\gamma_e=1.5$ is larger than that with $\gamma=2.0$, and the difference is resolvable by the instruments.
On the other hand, comparing the case of $\gamma_e=2.0$ with the case of $\gamma_e=2.5$, we find that the influence of a softer spectrum is not significantly. Because the majority of the halo's TeV emission is already emitted by recently injected electrons and the centroid is around the current position of pulsar. Although a softer injection spectrum would further reduce the contribution of relic electrons, it cannot make $\Theta$ significantly smaller.

Note that the situation is different at GeV band (shown in the top-left panel of Fig.~\ref{fig:thetaspe}), because a harder injection spectrum would lead to less electrons that radiate at such low energies, provided the same bolometric injection luminosity. Therefore, opposite to the case at TeV band, the separation angle at GeV would be reduced given a harder injection spectrum. However, since the cooling is not important at the low energy, the influence of the injection spectrum is not significant, except at 20\,GeV for $\Theta'$. This is because $\Theta'$ measures the separation between the brightest point from the pulsar in the intensity map . At GeV band, the morphology is double peaked due to the slow diffusion and the inefficient cooling of electrons, as discussed earlier. The intensity ratio between the peak at the pulsar's position and the peak at the original position increases with energy, whereas the position of the brightest point transfers from the original position at $E\lesssim 10\,$GeV to pulsar's current position at $E\gtrsim 10\,$GeV. Around 20\,GeV, the intensities between the two peaks are comparable to each other and hence the position of the brightest point in the intensity map becomes sensitive to model parameters even the influence is small.

We notice that the injection electron spectrum was assumed to be a broken power-law in some literature \citep[e.g.][]{Bucci11, Ishizaki17}, usually with a spectral break around $0.1-1$\,TeV. Since the spectrum after the break is soft, we do not expect that assuming a broken power-law spectrum has significant influence to the separation angle in the benchmark case.



\subsection{Dependence on the magnetic field strength}

\begin{figure*}[t]
\centering
\includegraphics[width=1.0\textwidth]{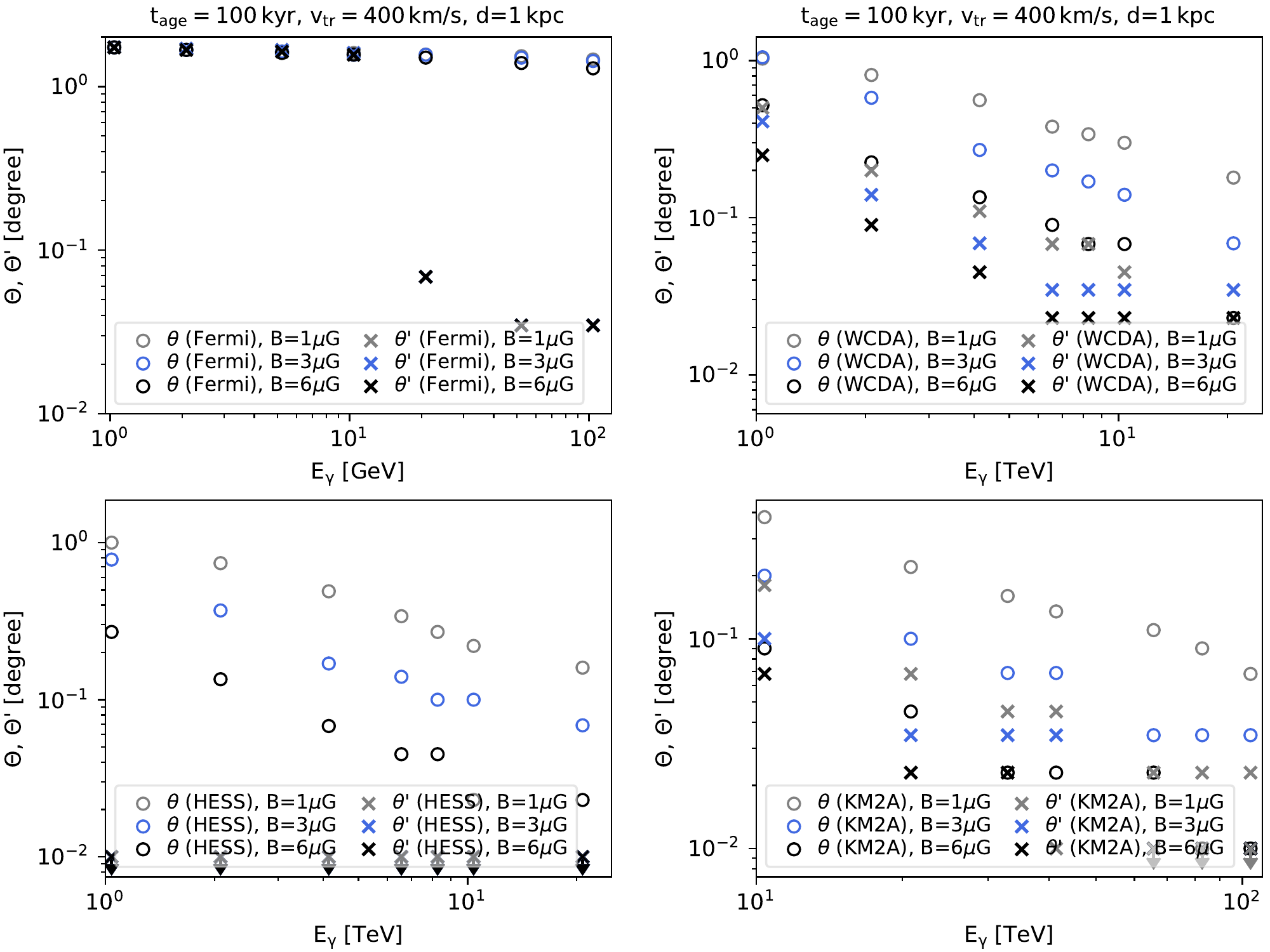}
\caption{The separation angle $\Theta$ and $\Theta'$ of the benchmark pulsar ($t_{\rm age}=$100\,kyr) but assuming different magnetic field $B=1,3,6\,\rm \mu G$ at different energies. }
\label{fig:theta_b}
\end{figure*}

The strengths of the radiation field and the magnetic field affect the cooling of electrons and, consequently, the separation angle of pulsar halo. The \gray emission we are concerned with is mainly produced by electrons up-scattering the FIR and CMB radiation. We find the influence of the radiation field on the separation angle is finite and the discussion is shown in Appendix A. On the other hand, the energy loss timescale through synchrotron process is comparable to the IC process and is the main energy loss process for $>10\,$TeV electrons when $B=3\,\mu G$. Thus the influence of $B$ is more important. We compare the separation angles of $B=1\,\mu G$, $B=3\,\mu G$ and $B_0=6\,\mu G$ in Fig.~\ref{fig:theta_b}. The separation angle of stronger magnetic field is smaller since the faster cooling of relic electrons and correspondingly less dominant \gray emission, vice versa. The difference of $\Theta$ is observable by LHAASO(WCDA) and HESS and reaches two times above 4\,TeV. In the interpretation of the observed offset of pulsar halo, the magnetic field strength can affect the theoretical calculation and conclusion.

\subsection{Evolution of separation angle with pulsar age}

\begin{figure}[h]
\centering
\includegraphics[width=0.9\columnwidth]{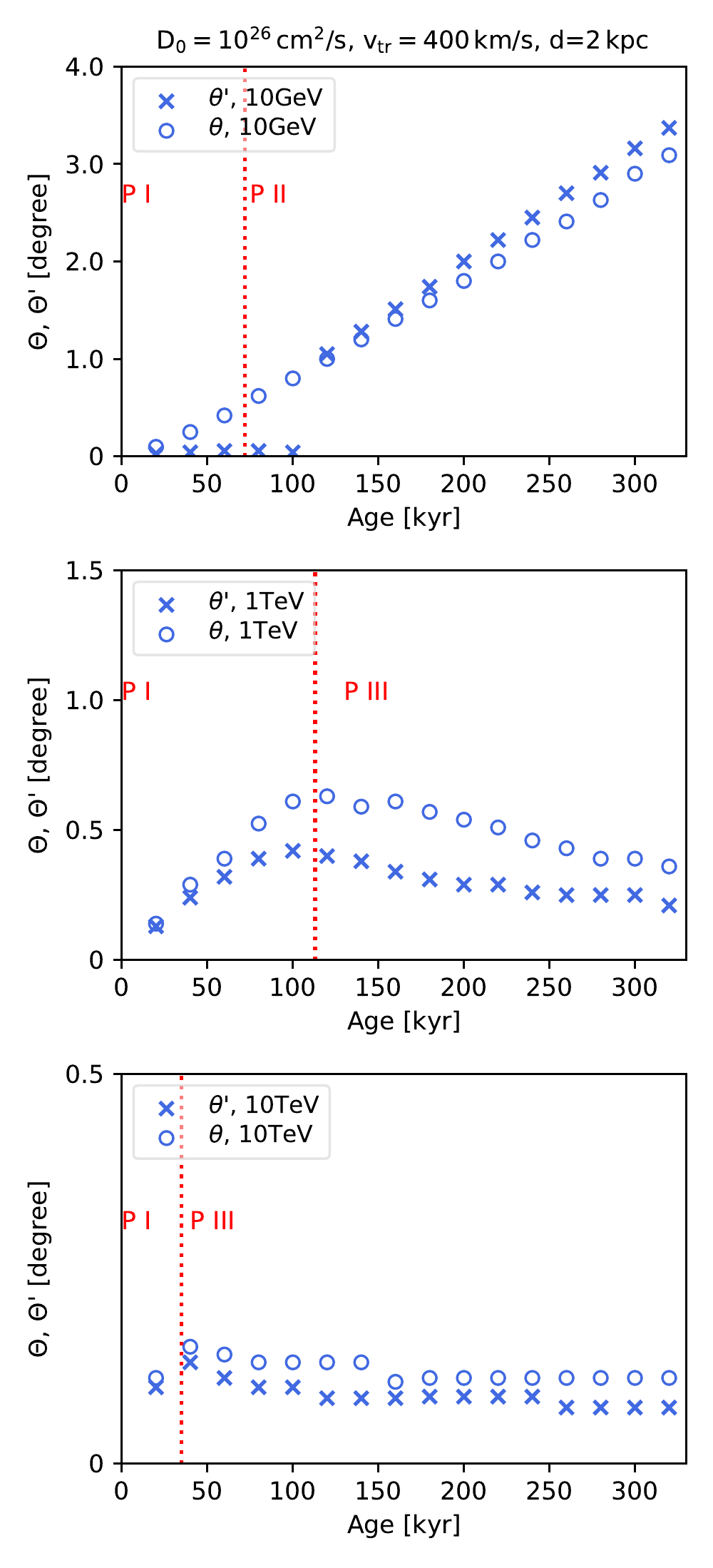}
\caption{The evolution of separation angle $\Theta$ and $\Theta'$ with $t_{\rm age}$ of benchmark pulsar, at 10\,GeV, 1\,TeV and 10\,TeV. }
\label{fig:thetatbmpf}
\end{figure}

The age of pulsar directly influences the displacement of pulsar and thus the separation angle of the pulsar halo. We calculate the separation angle at 10\,GeV, 1\,TeV and 10\,TeV of the benchmark pulsar with different $t_{\rm age}$. The result is plotted in Fig.~\ref{fig:thetatbmpf}, with evolutionary phases marked. At PHASE I and II $\Theta$ increases with $t_{\rm age}$ linearly and its value is limited by the displacement of pulsar. This reflects that the relic electrons contribute the majority of the \gray emission. When the pulsar halo enters PHASE III, $\Theta$ starts to decrease with $t_{\rm age}$. This is due to the cooling of relic electrons, the morphology of pulsar halo becomes more symmetric. For $\Theta'$, there is a jump after the pulsar halo entering PHASE II, as the pulsar halo transfers to double peak. And similar to $\Theta$, $\Theta'$ also decreases with $t_{\rm age}$ at PHASE III. Besides, the cooling timescale of tens TeV electrons is less than 50\,kyr, thus only electrons injected within this timescale can still radiate $>10\,$TeV \gray emission. Thus for the middle-aged pulsar ($t_{\rm age}>$100\,kyr), $\Theta$ and $\Theta'$ of $>10$\,TeV halo are nearly constant with $t_{\rm age}$.

\subsection{Maximum separation angle}\label{sec:max}
\begin{figure}[h]
\centering
\includegraphics[width=1.0\columnwidth]{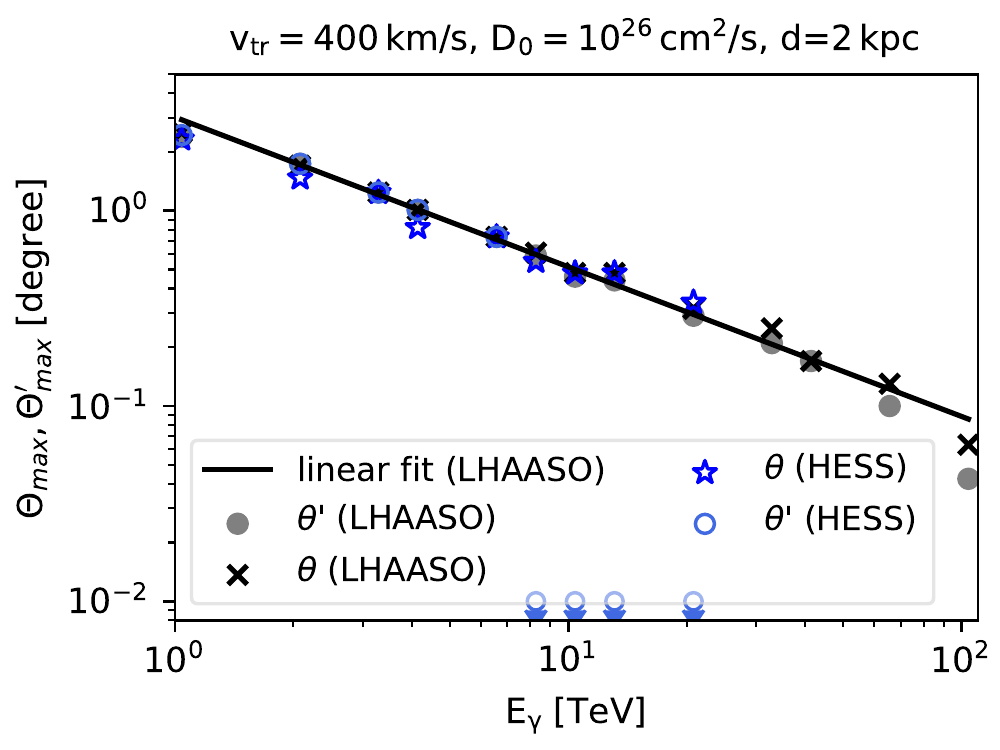}
\caption{The maximum separation angle $\Theta_{\rm max}$ and $\Theta'_{\rm max}$ at different energies that can be produced by a pulsar located at 2\,kpc with a proper velocity of $v_{\rm tr}$=400\,km/s.}
\label{thetainjs}
\end{figure}

The analytical estimation of the separation angles shown as black curves in Fig.~\ref{thetaE} largely exceeds the numerical results in PHASE \uppercase\expandafter{\romannumeral 3}. This is due to the ignorance of the continuous injection of electrons in the analytical formula as discussed above.
 We note that in certain condition the influence of the continuous injection could be suppressed. Practically, as the pulsar wind luminosity decreases with time, the maximum energy of electrons achievable in the termination shock may also decreases with time. When the maximum energy is below certain energy, we can regard the injection of electrons of this energy being ceased. The termination of injection would cause a larger offset because electrons are not injected at the position close to the present position of the pulsar. However, if the injection stops too early, i.e., the time experienced from the epoch of the injection termination to the present time, denoted by $t_{\rm s}$, is longer than the cooling timescale $t_{\rm c}$, we do not expect to see the emission of electrons at present. Therefore, we expect the most favorable condition for a large offset to be $t_{\rm s}\sim t_{\rm c}$. We then calculate the evolution of the pulsar halo at each energy with turning off the electron injection at $t=t_{\rm age}-t_{\rm c}$ to maximize the separation angle. The result is shown in Fig.~\ref{thetainjs}, assuming $B=3\,\rm \mu G$. The offset obtained in this condition ($\Theta_{\rm max}$) is independent on the electron injection history. We find that the $\Theta_{\rm max}-E_\gamma$ relation in the range of 1-100\,TeV can be empirically 
depicted by
\begin{equation}\label{eq:max_offset}
\Theta_{\rm max}=3^\circ \left(\frac{E_\gamma}{1\,\rm TeV}\right)^{-0.77}\left(\frac{v_{\rm tr}}{400\,\rm km/s}\right)\left(\frac{d}{2\,\rm kpc}\right)^{-1},
\end{equation}
except in the case that the offset is evaluated by $\Theta'$ whilst the halo is measured by an instrument of good angular resolution such as HESS.  
We note that the offset above 100\,TeV is quite small due to the rapid cooling of ultra-high energy electrons and is difficult to be resolved by the current \gray instrument, i.e., LHAASO-KM2A, unless the pulsar is located sufficiently close to Earth (e.g., at a distance of $\sim 100\,$pc from Earth) and has a high proper velocity (e.g., $\gtrsim 1000\,$km/s).

\section{Discussion}\label{sec:discussion}

\subsection{The evolution of spin-down luminosity of pulsar}

\begin{figure}[h]
\centering
\includegraphics[width=1.0\columnwidth]{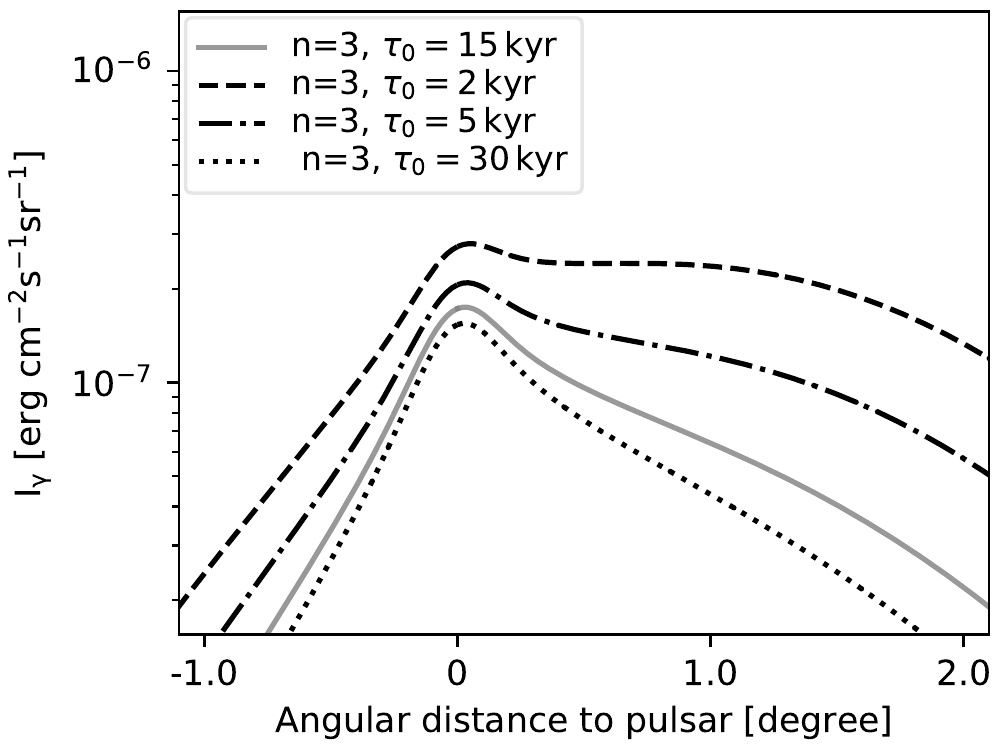}
\includegraphics[width=1.0\columnwidth]{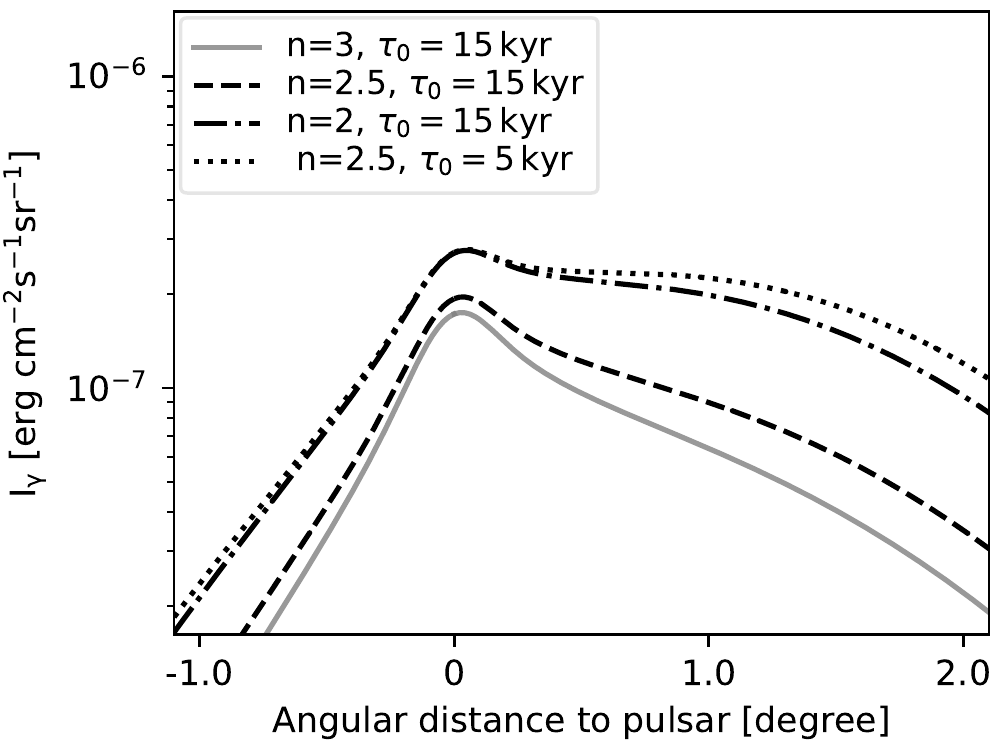}
\caption{The \gray intensity profile at 1\,TeV of benchmark pulsar ($t_{\rm age}=$100\,kyr), but with different evolution of $L_e(t)$, after convolving with the PSF of HESS. }
\label{fig:profilent}
\end{figure}

\begin{figure}[h]
\centering
\includegraphics[width=0.9\columnwidth]{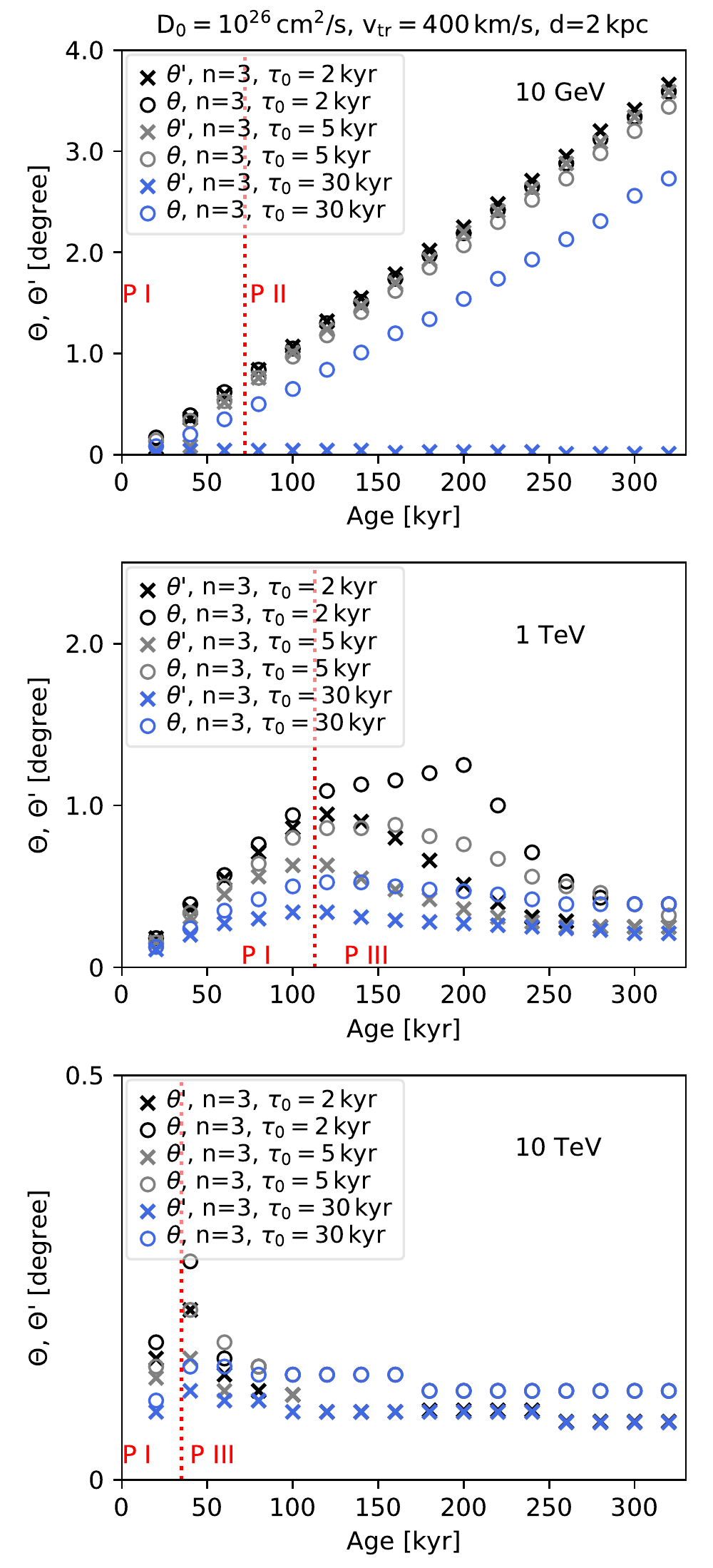}
\caption{The evolution of separation angle $\Theta$ and $\Theta'$ with $t_{\rm age}$ of pulsar with $n=3$ and $\tau_0=2,5,30\,\rm kyr$ respectively, at 10\,GeV, 1\,TeV and 10\,TeV. }
\label{fig:thetattau0}
\end{figure}

\begin{figure}[h]
\centering
\includegraphics[width=0.9\columnwidth]{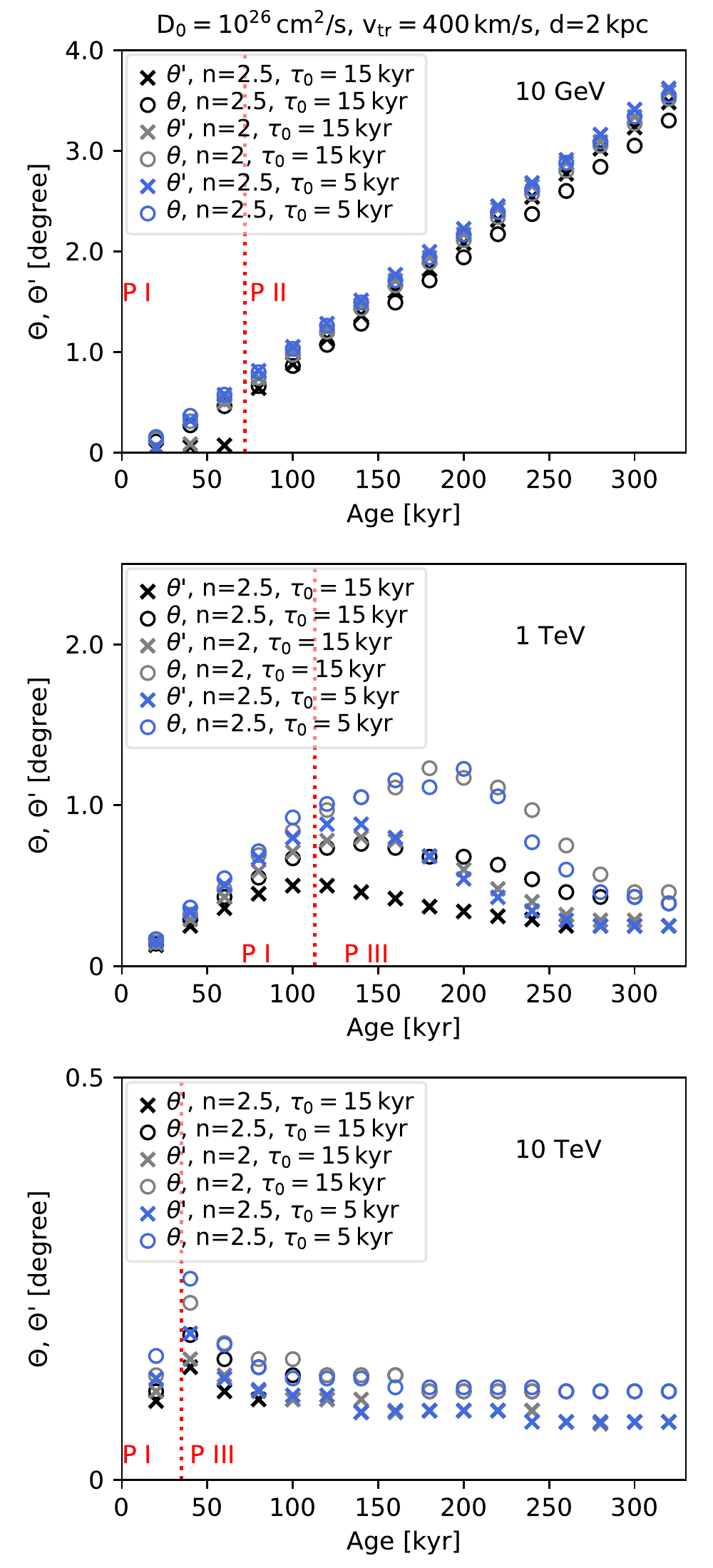}
\caption{The evolution of separation angle $\Theta$ and $\Theta'$ with $t_{\rm age}$ of pulsar with $n=2,2.5$ and $\tau_0=15\,\rm kyr$ respectively, $n=2.5$ and $\tau_0=5\,\rm kyr$, at 10\,GeV, 1\,TeV and 10\,TeV. }
\label{fig:thetatn}
\end{figure}

The electron luminosity injected by a pulsar is assumed to be proportional to the spin-down luminosity of pulsar, which is determined by the braking index $n$ and the initial spindown timescale $\tau_0$. In previous sections, we considered a typical value $n=3$, assuming the pulsar to be a magnetic dipole, and $\tau_0=15\,$kyr following the case of Geminga. However, these two parameters are not necessarily universal for other middled-aged pulsars. Thus,  we discuss the uncertainty in the electron injection history $L_e(t)$, i.e., $n$ and $\tau_0$, in this subsection.

The braking index $n$ is defined as $\ddot{v}v/\dot{v}^2$, where $v=2\pi/P$ is the angular frequency of pulsar and $\dot{v}(\ddot{v})$ is the time differential of $v(\dot{v})$. The direct measurement of $n$ requires the pulsar young enough to own a measurable $\ddot{v}$. Most of the young pulsars have $1<n<3$ \citep{Lyne93,Lyne96,Living05,Weltevrede11,Roy12,Hamil15}, but recently it is also found a pulsar with $n>3$ \citep{Archibald16}. On the other hand, these young pulsars may not represent the properties of older pulsars. The decay of pulsar magnetic field and inclination angle can induce the evolution of $n$ with time, i.e., increasing with time \citep{Tauris01,Johnston17}. Recently, \citet{Cholis18} used the measurement of electron and positron spectrum to constrain the spin-down properties of pulsars. They found that $n<2.5$ is disfavored for the majority of middle-aged pulsars ($>100\,\rm kyr$).

The initial spin-down timescale $\tau_0$ can be derived if the initial period of the pulsar, $P_0$, is further given. The measurement of $P_0$ is only possible for several pulsars associated SNR, so that one may estimate the true age of pulsar and derive $P_0$. Based on this method, \citet{Popov12} found that $P_0$ ranges from $20\,\rm ms$ to $300\,\rm ms$, in a Gaussian distribution with an average of $100\,\rm ms$ and standard deviation of $100\,\rm ms$. Another method to study the distribution of $P_0$ of Galactic pulsars is to search for a distribution of $P_0$ that can reproduce the properties of observed pulsar population, via modeling the evolution of pulsar from birth \citep{Faucher06,Johnston17}. Assuming $n$ being a constant, \citet{Faucher06} found that the distribution of $P_0$ has a mean value of $300\,\rm ms$ with a wide distribution. Combining above two methods, the derived $\tau_0$ of most middle-aged pulsars ranges from 2\,kyr to 100\,kyr. 

Besides, from the perspective of the selection effect, those middle-aged pulsars which can produce observable pulsar halos are less likely to have a small $n$ and/or a small $\tau_0$,  because such parameters lead to a rapid decline of the spindown luminosity of the pulsar and hence a lower electron injection luminosity at the present time.   



From a theoretical point of view, either a smaller $n$ or a smaller $\tau_0$ leads to a more rapid decline of the spindown luminosity with time, which enhances the relative contribution of the relic electrons, and vice versa.
To look into the influences of the electron injection history (i.e., $n$ and $\tau_0$), we compare the \gray profiles and separation angles under different combinations of the two parameters, including $\tau_0=2,5,30\,\rm kyr$ and $n=3$, $n=2,2.5$ and $\tau_0=15\,\rm kyr$, $n=2.5$ and $\tau_0=5\,\rm kyr$, while other parameters are the same with the benchmark parameters. We choose to show the corresponding profiles of the 1\,TeV \gray halo at $t_{\rm age}=100\,$kyr (Fig.~\ref{fig:profilent}), because the halo is at the transition stage from PHASE I to II/III (i.e., $t_{\rm age}\lesssim t_{c}\lesssim t_{\rm pd}$, where $t_{\rm c}=115\,$kyr, $t_{\rm pd}=126\,$kyr), and the morphological feature of the halo is sensitive to the electron injection history. We normalize the electron injection luminosity to be $L_e(t=100\,\rm kyr)=2.9\times10^{35}\,\rm erg/s$, which is the spindown luminosity of Geminga at 100\,kyr under the benchmark parameters. Note that although the luminosity at $t_{\rm age}$ is the same, the intensity at the current pulsar's position ($x=0$) is different. This is due to the different electron injection history.

The pulsar's present position is shifted by about 40\,pc from the initial position or $1.1^\circ$ for $d=2$\,kpc. Comparing to the solid grey curve, which represents the profile of the benchmark pulsar, the emission by relic electrons injected at early time leads to a flatter profile towards the pulsar's initial position with a smaller $n$ or $\tau_0$. Particularly, we see that for $n=3$ \& $\tau_0 = 2\,\rm kyr$ and $n=2.5$ \& $\tau_0=5\,$kyr, the \gray profile is nearly flat towards the direction of the pulsar's initial position, appearing the feature of PHASE II.
On the contrary, the halo with a larger $n$ or $\tau_0$ shows the features of pulsar halo at PHASE III, with a single peak at present position of pulsar. 

We also check the temporal evolution of the separation angle under different electron injection history for halos at 10\,GeV, 1\,TeV and 10\,TeV, as plotted in Fig.~\ref{fig:thetattau0} and \ref{fig:thetatn}. We can see that the separation angle is larger for a smaller $n$ and $\tau_0$, especially in the phase-transition stage. The reason is that the contribution of relic electrons is more important for a smaller $n$ and/or $\tau_0$, leading to a more asymmetric morphology of the halo. For example, when $\tau_0<5\,\rm kyr$ or $n<2.5$, the 1\,TeV \gray emission produced by relic electrons can maintain the dominance to the overall halo's emission even after the halo entering PHASE III, maintaining an increasing $\Theta$ until $t_{\rm age}\approx 2t_{\rm c}$. On the other hand, we also see that the separation angles under different combinations of $n$ and $\tau_0$ converge at $t_{\rm age}\gg t_{\rm c}$ or the deep PHASE III (e.g., at $t_{\rm age}=300\,$kyr, comparable to the age of Geminga). Comparing the evolution of the separation angle at 1\,TeV and that at 10\,TeV, we see that the higher the $\gamma$-ray energy, the faster it converges. Indeed, when relic electrons have been severely cooled in the deep PHASE III, the injection history can barely influence the present morphology of the halo. 

\subsection{Application to the observation}

\begin{table*}[ht]
\centering
\caption{3HWC and LHAASO sources with TeV halo candidate pulsars}\label{3HWC}
\begin{tabular}{ccccccc}
\hline\hline
3HWC& Pulsar& $\tau_{c}$(kyr)&d (kpc)&$v_{\rm tr}$(km/s)&$\theta_{\rm obs}$($^\circ$)&Comment\\ 
\hline
J0540+228&B0540+23&253&1.56&215&0.83&B$<1\,\rm \mu G$ or $n<2$\\
J0543+231&B0540+23&253&1.56&215&0.36&Unaligned\\
J0631+169&J0633+1746&342&0.19&128&0.95&Possible\\
J0634+180&J0633+1746&342&0.19&128&0.38&Unaligned\\
J0659+147&B0656+14&111&0.29&60&0.51&Unaligned\\
J0702+147&B0656+14&111&0.29&60&0.77&Unaligned\\
J1739+099&J1740+1000&114&1.23&-&0.13&Unclear\\
J1831-095&J1831-0952&128&3.68&-&0.27&Unclear\\
J1912+103&J1913+1011&169&4.61&-&0.31&Unclear\\
J1923+169&J1925+1720&115&5.06&-&0.67&Unclear\\
J1928+178&J1925+1720&115&5.06&-&0.85&Unclear\\
J2031+415&J2032+4127&201&1.33&-&0.11&Unclear\\
\hline
\hline
LHAASO& Pulsar& $\tau_{c}$(kyr)&d (kpc)&$v_{\rm tr}$(km/s)&$\theta_{\rm obs}$($^\circ$)&Comment\\
\hline
J2032+4102&J2032+4127&201&1.4\tablenotemark{a}&20.4\tablenotemark{b}&0.42& unlikely\\
J1929+1745&J1928+1746&82.6&4.6&-&0.25&$v_{\rm tr}>2700\,\rm km/s$\\
\hline
\end{tabular}
\tablecomments{
\flushleft{Pulsar halo candidates in 3HWC and LHAASO source list (column 1), related pulsars (column 2), the pulsar's characteristic age (column 3), distance (column 4), velocity (column 5) and observed separation between pulsar and source (column 6). The properties of pulsars are from the ATNF pulsar catalog \citep{Manchester05} or otherwise specified. Our brief comment on whether the separation between the pulsar and the source can be explained by pulsar proper motion is given in the last column:\\
'Unaligned' means the association is impossible due to the unaligned pulsar motion direction with the relative position of 3HWC source and candidate pulsar;\\
'Possible' means the offset can in principle be explained by the proper motion of the pulsar, given the measured the velocity and direction of the pulsar's proper motion;\\
'Unclear' means no measurement on the proper motion of the pulsar, but the offset may be explained by the proper motion if both the velocity and the direction are appropriate.\\
$^{a}$ \citet{Rygl12}\\
$^{b}$ \citet{Jennings18}\\
}
}
\end{table*}

As we discussed above, it is increasingly difficult to produce resolvable offset between the center of the pulsar halo and the pulsar by the proper motion at higher energy. An observable $\Theta$ requires that either the pulsar is located close-by, its proper velocity is extremely high, the magnetic field is weak or its braking index and $\tau_0$ of pulsar is small. However, offsets at such a high energy between extended TeV sources and the positions of associated middle-aged pulsars have already been observed by HAWC and LHAASO.

The 3HWC catalogue gives 12 extended TeV sources and the separation between the sources and their candidate pulsars \citep{HAWC20_3HWC}. Note that the analysis of the 3HWC catalogue does not optimize the sizes of the sources and most sources are identified as point sources, so the reported source position is not necessary to reveal the true centroid of the source. Nevertheless, we compare the theoretical separation angle induced by pulsar motion and the observed offsets to roughly judge if the association is possible. We list the 3HWC sources and potential pulsars in Table.~\ref{3HWC}. If we assume the separation is caused by pulsar proper motion, four 3HWC sources can be excluded from the association with candidate pulsars due to the not aligned pulsar motion direction. To associate 3HWC J0540+228 with pulsar B0540+23, the magnetic field of the ISM needs to be smaller than $1\,\rm \mu G$, or the braking index of pulsar is at least smaller than 2. For other eight 3HWC sources, the association is possible with proper combination of parameters and we need further knowledge of the proper motion velocity of their candidate pulsars to give a robust judgement.

Very recently, \cite{LHAASO21} reported discovery of 12 \gray sources above 100 TeV with more than $7\sigma$ statistical significance. Among them, two pulsar halo candidates, i.e. LHAASO J2032+4102 and LHAASO J1929+1745, are possibly associated with middle-aged pulsar PSR~J2032+4127 and PSR~J1928+1746 respectively. We may compare the observed separation with the theoretical maximum $\Theta$ induced by pulsar proper motion (Eq.\ref{eq:max_offset}) to quickly estimate whether the association could be true. We find that LHAASO J2032+4102 is impossible to be entirely associated with PSR J2032+4127 due to the small $v_{\rm tr}$ of pulsar, and the association of LHAASO J1929+1745 with PSR J1928+1746 requires an extremely large $v_{\rm tr}$. The results are also shown in Table.~\ref{3HWC}.
 
If a 3HWC source or a LHAASO source is truly a pulsar halo but the spatial offset between the source and the related pulsar is very large, we have to resort to more complicated but probably realistic scenarios, such as the anisotropic particle diffusion scenario if the geometric configuration of the magnetic field in the surrounding ISM of the pulsars is not chaotic \citep{Liu19_prl}. Let us envisage a specific scenario that the magnetic field in the east side of a pulsar is largely parallel to the periphery of the PWN while in the west side is largely radial. The diffusion of escaping electrons towards east is then suppressed because the cross-field diffusion is slow, while the electrons can quickly diffuse towards west. The pulsar halo under such a configuration of magnetic field would be significantly extended towards the west side and a large offset is expected between the pulsar and the centroid of the halo. Of course, the source is not necessarily a pulsar halo. Some sources could have an asymmetric morphology intrinsically. With certain specific conditions, a $\sim 10\,$kyr-aged PWN could vigorously expand \citep{Khangulyan18} while the asymmetric reverse shock arising from the supernova ejecta could crush one side of the PWN \citep{Blondin01,Gaensler03, HESS06_J1825}, and/or electrons preferentially escape from one side of the PWN due to the geometry of the magnetic field in the PWN \citep{Liu20}. These scenarios might cause a significant offset between the centroid of the TeV emission and the pulsar's location. Alternatively, the source could be composed of multiple origins. There are more than one source candidate around the two LHAASO sources, as shown in Extended Data Table 2 in \citet{LHAASO21}. \citet{Giacinti20} suggested that the energy density of the relativistic electron density inside the source being smaller than that of the ISM might be a criteria for the pulsar halo. Multiwavelength observations with high angular resolution will be crucial to reveal the true nature of the sources.

\subsection{Influence of two diffusion zones}

In our calculation of electrons diffusion, we consider a spatially homogeneous diffusion coefficient around the pulsar at a scale of $\sim 100\,$pc. 
Some previous studies suggest that there could exist two diffusion zones in this region, with a slow diffusion zone within a few tens of parsecs around the pulsar and a normal diffusion zone beyond the radius \citep{Fang18, Profumo18, Tang19}. Our present numerical treatment cannot deal with two-zone diffusion with the proper motion of pulsar, because the diffusion would become highly anisotropic if the pulsar has moved close to the boundary between the slow diffusion zone and the normal diffusion zone. Such a situation could happen since the displacement of the pulsar due to proper motion $v_{\rm p}t_{\rm age}$ could be comparable to the size of the slow diffusion zone.
Nevertheless, we may infer the influence of two-zone diffusion qualitatively. 
If the slow diffusion zone is centered at the initial position of pulsar, a large fraction of the huge amount of electrons injected at early epoch would be still confined inside the slow diffusion zone if not cooled. On the other hand, the electrons injected recently would more likely diffuse in the normal diffusion zone, and leads to a comparatively low particle density around the current position of the pulsar. As a consequence, it is beneficial to yield a large offset. 
On the contrary, if the slow diffusion zone moves with the pulsar, or it is centered at the pulsar's present position, electrons injected at early epoch would largely diffuse in the normal diffusion zone and hence their radiation becomes diffuse and faint, while electrons injected recently would diffuse in the slow diffusion zone and form a comparatively high density around the pulsar. As a result, the halo would be dominated by the radiation of electrons injected recently, appearing more symmetrical with a smaller offset. We leave the detailed and quantitative discussion of different mechanisms of two-zone diffusion and their influences in the future work. Note that the two-zone diffusion model mainly affects the morphology of pulsar halo at GeV-TeV, which can extend beyond the size of the slow diffusion zone. For the pulsar halo above 10\,TeV (PHASE III), the situation is similar to the single diffusion zone scenario, because the size of the halo is generally limited due to rapid cooling.

Our analysis here is consistent with the result of \citet{Johan19}. They considered the two-zone diffusion model to explain the pulsar halo of Geminga. They considered three scenarios about the slow diffusion zone: (1) the slow diffusion zone centers at the birthplace of the pulsar; (2) it centers at the current position of the pulsar; (3) it moves with the pulsar. For the first scenario, to be compatible with the observation, the slow diffusion zone of the second scenario needs to be large enough to include both the birthplace and the current position of Geminga, which is similar to the one-zone scenario. The other two scenarios result in more symmetrical halo, with a single peak and a short tail, due to the reason we discussed above.

\section{Conclusions}

In this paper, we studied the theoretical \gray morphology of a pulsar halo and the spatial offset between the halo and its associated pulsar considering the pulsar's proper motion.  We divide the evolution of the pulsar halo into three characteristic phases, based on three timescales, namely, when the displacement of the pulsar equal to the diffusion length of the initially injected particle $t_{\rm pd}$, the cooling timescale of electrons $t_{\rm c}$, and the age of the pulsar $t_{\rm age}$. For pulsar with $t_{\rm age}<t_{\rm pd}$ and $t_{\rm age}<t_{\rm c}$, the pulsar halo would appear a single-peak morphology with a broad peak (PHASE \uppercase\expandafter{\romannumeral 1}). For $t_{\rm pd}<t_{\rm age}<t_{\rm c}$, the morphology of the pulsar halo would show significant asymmetry with two humps or one hump with an extended plateau/tail (PHASE \uppercase\expandafter{\romannumeral 2}). For $t_{\rm c}<t_{\rm age}$, the morphology becomes single peaked again but with a narrow peak centred at the current position of the pulsar (PHASE \uppercase\expandafter{\romannumeral 3}). Note that in the phase-transition stage (i.e., $t_{\rm pd}\sim t_{\rm age}$, or $t_{c}\sim t_{\rm age}$), the morphological feature of a phase may not be distinct and is affected by the electron injection history. Our study can give a clue to the origin of the asymmetrical morphology of the pulsar halos at GeV -- TeV band.  If the age and the proper velocity of a pulsar is known, we can quickly figure out which phase its pulsar halo is experiencing and judge whether the observed morphology is consistent with the pulsar halo origin. This would be helpful to understand the origin of the extended gamma-ray sources. 

We defined two kinds of offsets between the pulsar halo and the pulsar: one is the separation between the centroid of the halo and the pulsar's position ($\Theta$) and the other is the separation between the brightest position in the halo and the pulsar's position ($\Theta'$). The influence of various model parameters on the separation angles are studied. We found that both separation angles generally decrease with increasing \gray energy. The offset become difficult to be resolved above 10\,TeV due to the very rapid cooling of the emitting electrons, unless the pulsar is in close proximity to Earth (e.g., $\lesssim 100$\,pc) and/or has a high proper velocity ($>1000\,$km/s). We also found that $\Theta$ and $\Theta'$ have weak dependence on some parameters, i.e., the electron injection spectrum and background photon field. Other parameters, such as the magnetic field strength of ISM, affect $\Theta$ significantly and should be treated carefully when interpreting the observation. From the dependence we obtained, we can estimate the theoretical separation angle. Then, by comparing it to the observed value, we can either exclude the possibility of some association between the candidate pulsar and extended emission observed by HAWC or LHAASO, or put constraints on parameters that needed to make the association possible.

Under an extreme assumption that electron injection terminates at $t=t_{\rm age}-t_{\rm c}$, we gave the maximum separation angle ($\Theta_{\rm max}$ and $\Theta'_{\rm max}$) that the pulsar's proper motion can induce at PHASE III. We found that the maximum separation angle of the emission centroid can be given by $\Theta_{\rm max}=3^\circ (E_\gamma/1\,{\rm TeV})^{-0.77}(v_{\rm tr}/400\,{\rm km~s^{-1}})(d/2\,\rm kpc)^{-1}$ empirically,  assuming $B=3\,\rm \mu G$. Even for instrument of good angular resolution such as HESS, we do not expect to see offset between the brightest point in the halo and the position of the pulsar for reasonable parameters. Therefore, if too large an offset is detected, it would require the consideration of a more complex configuration of the magnetic field and correspondingly anisotropic diffusion of electrons. Alternatively, it might imply that the source is not simply a pulsar halo. Multi-wavelength observation with high angular resolution would then be helpful to reveal the true origin of the \gray source. So far, there is only few pulsar halos detected. Future observation by LHAASO and HAWC should be able to bring us more samples and provide the opportunity to explore the mechanism of pulsar halo offset and test our model.

\section*{Acknowledgments}
We thank the anonymous referee for the constructive comments and suggestions that help us to improve the quality of the paper. This work is supported by the National Key R \& D program of China under the grant 2018YFA0404203 and the NSFC  grants
11625312, 11851304 and U2031105.

\bibliographystyle{apj}
\bibliography{ms}

\section*{Appendix A: Dependence of separation angle on radiation field}

\begin{figure*}[ht]
\centering
\includegraphics[width=1.0\textwidth]{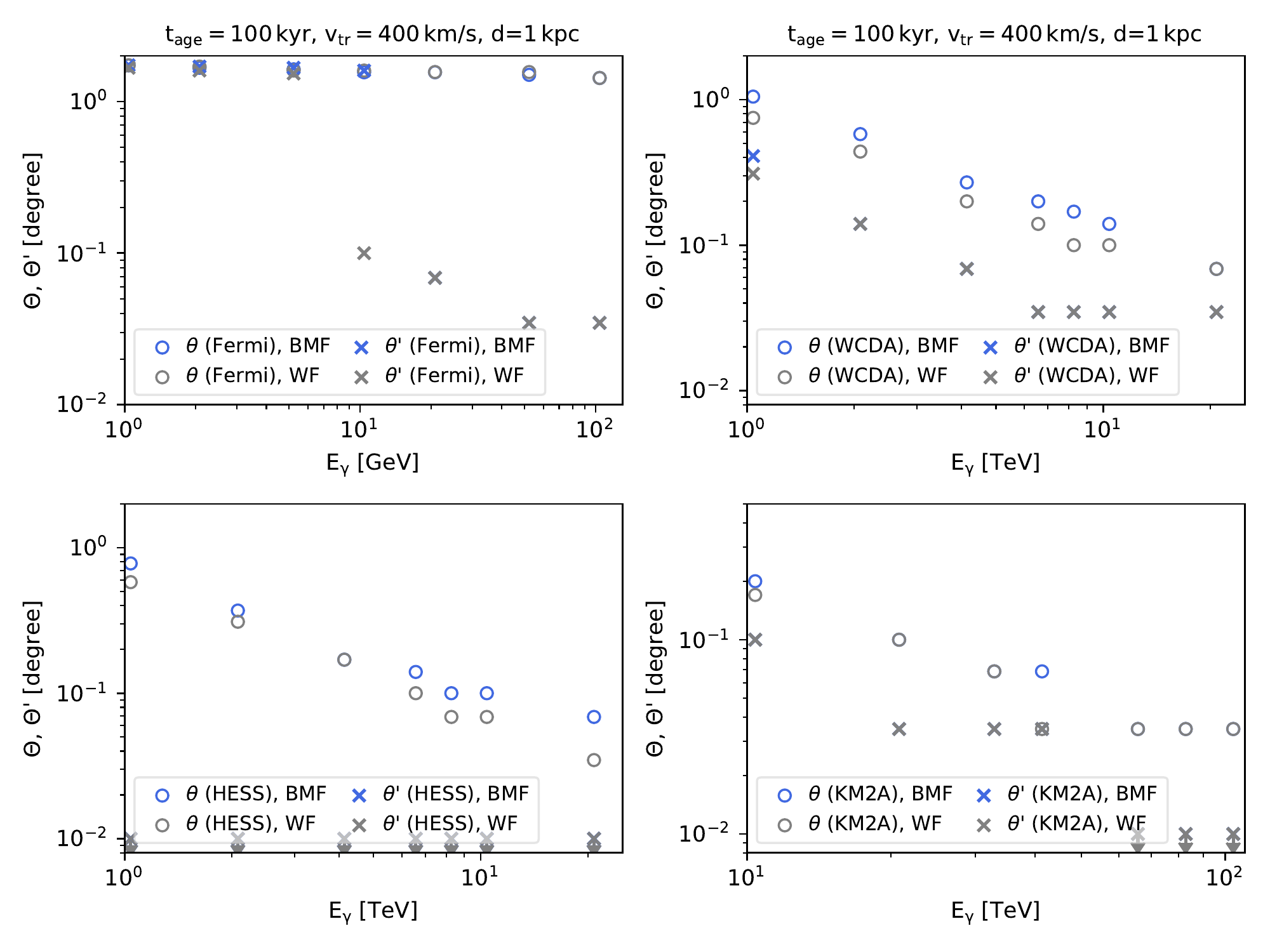}
\caption{The separation angle $\Theta$ and $\Theta'$ of the benchmark pulsar ($t_{\rm age}=$100\,kyr) but assuming different background radiation field at different energies. The blue points are assuming the benchmark radiation field (BMF) and the gray points are assuming the weaker radiation field (WF).}
\label{fig:theta_wpf100}
\end{figure*}

As we discussed in the main text, the strength of radiation field affects the cooling of electrons and correspondingly affects the separation angle of pulsar halo. Here we take a weaker radiation field (the one at 13\,kpc from galactic center) around the pulsar as: CMB ($T=2.73\,\rm K$ and $U=0.25\,\rm eVcm^{-3}$), far-infrared radiation (FIR) field ($T=30\,\rm K$ and $U=0.1\,\rm eVcm^{-3}$), near-infrared radiation field ($T=500\,\rm K$ and $U=0.04\,\rm eVcm^{-3}$) and visible light radiation field ($T=5000\,\rm K$ and $U=1.9\,\rm eVcm^{-3}$) and compare the separation angles of the benchmark model photon field used in the main text (BMF) with those of this weaker photon field (WF). The result is plotted in Fig.~\ref{fig:theta_wpf100}.  Overall, the separation angle of weaker photon field is smaller than the one of benchmark photon field. This is because in the benchmark photon field, the gamma-ray emission from GeV to about ten TeV are produced by electrons and FIR field. But with 10 times weaker FIR field, the GeV-TeV gamma-ray emission is produced by the IC process between electrons and CMB field in the WF model. Thus higher energy electrons are needed in WF model and correspondingly shorter cooling timescale causes the gamma-ray emission is more focused around the present pulsar position. The difference of $\Theta$ induced by different radiation field is observable only around 1\,TeV by HESS, we thus conclude the influence of different photon field around pulsar is finite. 

\end{document}